\documentclass[twocolumn]{aastex6}
\pdfoutput=1

\usepackage{amsmath}
\usepackage{microtype}
\renewcommand{\vec}[1]{\mathbf{#1}}
\newcommand{\dd}{\,\mathrm{d}}
\newcommand{\pd}{\partial}

\begin{document}

\title{Fully kinetic versus reduced-kinetic modelling of collisionless plasma turbulence}

\author{
Daniel~Gro\v selj\altaffilmark{1,2},
Silvio~S.~Cerri\altaffilmark{3},
Alejandro~Ba\~ n\' on Navarro\altaffilmark{1,2},
Christopher~Willmott\altaffilmark{4},
Daniel~Told\altaffilmark{1}, 
Nuno~F.~Loureiro\altaffilmark{4},
Francesco Califano\altaffilmark{3}, and
Frank~Jenko\altaffilmark{1,2}
}

\altaffiltext{1}{Max-Planck-Institut f\" ur Plasmaphysik, Boltzmannstra\ss e 2, 
85748 Garching, Germany}

\altaffiltext{2}{Department of Physics and Astronomy, University of California, 
Los Angeles, CA 90095, USA}

\altaffiltext{3}{Physics Department ``E.~Fermi'', University of Pisa, Largo B.~Pontecorvo 3,
56127 Pisa, Italy}

\altaffiltext{4}{Plasma Science and Fusion Center, Massachusetts Institute of Technology, 
Cambridge, MA 02139, USA}

\begin{abstract}
We report the results of a direct comparison between different kinetic models of 
collisionless plasma turbulence in two spatial dimensions. The models considered 
include a first principles fully kinetic (FK) description, two
widely used reduced models [gyrokinetic (GK) and hybrid-kinetic (HK)
with fluid electrons], and a novel reduced gyrokinetic approach (KREHM). Two
different ion beta ($\beta_i$) regimes are considered: 0.1 and 0.5. 
For $\beta_i=0.5$, good agreement between the GK and FK models is found at 
scales ranging from the ion to the electron gyroradius, thus providing firm 
evidence for a kinetic Alfv{\' e}n cascade scenario.
In the same range, the HK model produces shallower spectral slopes, presumably 
due to the lack of electron Landau damping. For $\beta_i=0.1$, a detailed
analysis of spectral ratios reveals a slight disagreement between 
the GK and FK descriptions at kinetic scales, even though kinetic Alfv{\' e}n 
fluctuations likely still play a significant role. 
The discrepancy can be traced back to scales above the ion gyroradius, where 
the FK and HK results seem to suggest the presence of 
fast magnetosonic and ion Bernstein modes in both plasma beta regimes, but 
with a more notable deviation from GK in the low-beta case. 
The identified practical limits and strengths of reduced-kinetic approximations, 
compared here against the fully kinetic model on a case-by-case basis, 
may provide valuable insight into the main kinetic effects at play in 
turbulent collisionless plasmas, such as the solar wind.
\end{abstract}


\section{Introduction}

Turbulence is pervasive in astrophysical and space plasma environments, 
posing formidable challenges for the explanation of complex phenomena 
such as the turbulent heating of the solar wind~\citep{bruno2013}, 
magnetic field amplification by dynamo action~\citep{Kulsrud2008, Rincon2016, Kunz2016},
and turbulent magnetic reconnection~\citep{Matthaeus1986, Lazarian1999, Loureiro2009, Matthaeus2011, Lazarian2015, Cerri2017a}. 
The average mean free path of the ionized particles in the majority of these naturally turbulent systems 
typically greatly exceeds the characteristic scales of turbulence, rendering the dynamics over 
a broad range of scales essentially collisionless. Driven by increasingly 
accurate in-situ observations, a considerable amount of current research is focused 
on understanding collisionless plasma turbulence at kinetic scales of 
the solar wind~\citep{Marsch2006, bruno2013}, where the phenomenology 
of the inertial range (magnetohydrodynamic) turbulent cascade is no longer applicable
due to collisionless damping of electromagnetic fluctuations, dispersive 
properties of the wave physics, and as a result of 
kinetic-scale coherent structure formation~\citep{Kiyani2015}. 
The research effort with emphasis on kinetic-scale turbulence includes 
abundant observations \citep{Leamon1998, Bale2005, Alexandrova2009, Sahraoui2009, 
Kiyani2009, Osman2011, He2012, Salem2012, Chen2013, Lacombe2014, Chasapis2015, Perrone2016, Narita2016}, 
supported by analytical predictions \citep{Galtier2003, Howes2008, Schekochihin2009, Boldyrev2013, Passot2015}, 
and computational studies \citep{Howes2008a, Saito2008, Servidio2012, Verscharen2012, Boldyrev2012, 
Wu2013a, Karimabadi2013a, TenBarge2013, Chang2014, Vasquez2014, Valentini2014, Told2015, 
Wan2015, Franci2015, Cerri2016, Parashar2016}.

Out of numerical convenience and based on physical considerations, the simulations of collisionless plasma turbulence 
are often performed using various types of simplifications of the first principles fully kinetic 
plasma description. Among the most prominent reduced-kinetic 
models for astrophysical and space plasma turbulence are the so-called 
hybrid-kinetic (HK) model with fluid electrons \citep{Winske2003, Verscharen2012, 
Parashar2009, Vasquez2014, Servidio2015, 
Franci2015, Kunz2016} and the gyrokinetic (GK) model \citep{Howes2008, Howes2008a,
Schekochihin2009, TenBarge2013, Told2015, Li2016}. Even though these models 
are frequently employed, no general consensus concerning 
the validity of reduced-kinetic treatments presently exists within the community, 
with distinct views on the subject being favoured by 
different authors \citep{Howes2008, Matthaeus2008, Schekochihin2009, Servidio2015}. 
Fully kinetic (FK) simulations, on the other hand, are 
currently constrained by computational requirements that limit the prevalence of such studies, the 
accessible range of plasma parameters, and the ability to extract relevant information from the 
vast amounts of simulation data \citep{Saito2008, Wu2013a, Karimabadi2013a, Chang2014, Haynes2014, Wan2015, Parashar2016}. 
In order to spend computational resources wisely and focus fully kinetic simulations 
on cases where they are most needed, it is therefore crucial 
to establish a firm understanding of the practical limits of reduced-kinetic models. 
Furthermore, the question concerning the validity of reduced-kinetic approximations is not only 
important from the computational perspective, but is also of major interest for the development 
of analytical predictions based on a set of simplifying but well-validated assumptions, 
and for the interpretation of increasingly accurate in-situ satellite measurements. 
Last but not least, a thorough understanding of the range of parameters and 
scales where reduced models correctly reproduce the fully kinetic results is instrumental in 
identifying the essential physical processes that govern turbulence in those regimes.

In this work, we try to shed light on some of the above-mentioned issues by performing a systematic, 
first-of-a-kind comparison of the FK, HK, and GK modelling approaches for the particular case of collisionless, 
kinetic-scale plasma turbulence. Furthermore, we also compare the results against a novel reduced gyrokinetic model, 
corresponding to the low plasma beta limit of GK. Studies of similar type have 
been relatively rare so far \citep{Birn2001, Henri2013, TenBarge2014a, Munoz2015,
Told2016, Camporeale2017, Cerri2017b, Pezzi2017a, Pezzi2017}. Considering these circumstances, we choose here to investigate
different plasma regimes in order to give a reasonably comprehensive view of the effects 
of adopting reduced-kinetic approximations for studying kinetic-scale, collisionless plasma turbulence.
On the other hand, the simulation parameters that we select per force result 
from a compromise between the desire for maximal relevance of results in relation to solar wind turbulence 
and computational accessibility of the problem. To this end, we adopt a two-dimensional and widely-used 
setup of decaying plasma turbulence \citep{Orszag1979} and carefully tailor the simulation 
parameters to reach---as much as possible---the plasma conditions relevant for the solar wind. The particular 
type of initial condition was chosen here for its simplicity and popularity, allowing for a relatively 
straightforward reproduction of our simulations and/or comparison of the results 
with previous works \citep{Biskamp1989, Dahlburg1989, Politano1989, Politano1995, 
Parashar2009, Parashar2015, Loureiro2016, Li2016}. On the other hand, 
it is still reasonable to expect that the turbulent system 
studied in this work shares at least some qualitative similarities 
with natural turbulence occurring at kinetic 
scales of the solar wind \citep{Servidio2015, Li2016, Wan2016}.
We also note that the choice of initial conditions should not affect the 
small-scale properties of turbulence due to the self-consistent reprocess of 
the fluctuations by each model during the turbulent cascade~\citep{Cerri2017b}. 

The paper is organized as follows. In Section~\ref{sec:models} we provide a short summary of the kinetic models included in
the comparison, followed by a description of the simulation setup in Sec.~\ref{sec:desc}. 
The results presented in Sec.~\ref{sec:results} are divided 
into several parts. In the first part, we investigate the spatial structure of the solutions by comparing the snapshots 
of the electric current and the statistics of magnetic field increments (Sec.~\ref{sec:spatial_structure}). 
The analysis of turbulent structures in real space is followed by a comparison of the global turbulence 
energy budgets for each model (Sec.~\ref{sec:energy}). 
Afterwards, we perform a detailed comparison of the spectral properties of the solutions (Sec.~\ref{sec:spectra}), followed
by the analysis of the nonthermal ion and electron free energy fluctuations in the spatial domain (Sec.~\ref{sec:free_energy}). 
Finally, we conclude the paper with a summary and discussion of our main results (Sec.~\ref{sec:conclusions}).

\section{Kinetic Models Included in the Comparison}
\label{sec:models}

For the sake of clarity, we provide below brief descriptions of the kinetic models involved in the comparison. 
However, we make no attempt to give a comprehensive overview of each model and we refer the reader to the 
references given below for further details. 
As is often done in literature, we adopt here the so-called collisionless approximation \citep{Lifshitz1981, Klimontovich1997} 
as the basis for interpreting our results, given the fact that the solar wind is very weakly collisional \citep{Marsch2006, bruno2013}. 
By definition, the collisionless approximation neglects any discrete binary (and higher order) particle interactions. 
It is worth mentioning that although we do not explicitly consider the role of a weak collisionality, 
our numerical solutions still contain features which resemble (weakly) collisional effects in a plasma. In the particle-in-cell (PIC) method, 
collisional-like effects occur due to random statistical fluctuations in the electrostatic potential 
felt by each finite-size particle, and due to numerical effects 
related to the use of a spatial grid for the electromagnetic fields \citep{Okuda1970, Hockney1971, Birdsall2005}. 
Eulerian based methods, on the other hand, are subject to an effective velocity space diffusion, 
which acts to smooth the particle distribution function at the smallest 
resolved velocity scales.

We start with a quick summary of the fully kinetic, 
electromagnetic plasma model \citep{Klimontovich1967, Lifshitz1981, Klimontovich1997, Liboff2003}. 
In the collisionless limit, the kinetic properties of a plasma are governed by the 
Vlasov equation for each particle species $s$:
\begin{align}
\frac{\pd f_s}{\pd t} + \vec v\cdot\nabla f_s + 
q_s\left(\vec E + \frac{\vec v\times\vec B}{c}\right)\cdot\frac{\pd f_s}{\pd\vec p} = 0,
\end{align}
where $f_s(\vec r,\vec p, t)$ is the single-particle distribution function, $q_s$ is the species charge, 
$\vec v$ is the velocity, $\vec p$ is the (relativistic) momentum, 
and $\vec E$ and $\vec B$ are the (smooth) self-consistent electromagnetic fields. 
The self-consistent fields are obtained from the full set of Maxwell equations,
\begin{align}
\frac{\pd\vec E}{\pd t}& = c\,\nabla\times\vec B - 4\pi\vec J,&
\frac{\pd\vec B}{\pd t}& = -\, c\,\nabla\times\vec E,\label{eq:me1}\\
\nabla\cdot\vec E& = 4\pi\rho, & \nabla\cdot\vec B& = 0,\label{eq:me2}
\end{align}
where $\vec J=\sum_s q_s\int\vec v f_s\dd^3{\vec p}$ is the electric current 
and $\rho = \sum_s q_s\int f_s\dd^3\vec p$ is the charge density. The 
above equations form a closed set for the fully kinetic description 
of a plasma in the collisionless limit.

The hybrid-kinetic Vlasov-Maxwell model \citep{Byers1978, Harned1982, Winske2003} 
is a non-relativistic, quasi-neutral plasma model 
where the ions are fully kinetic and the electrons are treated as 
a background neutralizing fluid with some underlying assumption for their equation of state 
(typically an isothermal closure). 
The fundamental kinetic equation solved by the HK model is the Vlasov equation for the ions:
\begin{align}
\frac{\pd f_i}{\pd t} + \vec v\cdot\nabla f_i + \frac{q_i}{m_i}\left(\vec E + \frac{\vec v\times\vec B}{c}\right)\cdot\frac{\pd f_i}{\pd\vec v} = 0,
\end{align}
where $f_i(\vec r,\vec v)$ is the ion single-particle distribution function and 
$m_i$ is the ion mass. The $\vec B$ field is advanced using Faraday's law
\begin{align}
\frac{\pd\vec B}{\pd t} & = -\, c\,\nabla\times\vec E.
\end{align}
Since the HK model assumes the non-relativistic limit,
the displacement current is neglected in the Amp\'ere's law, which thus reads
\begin{align}
\nabla\times\vec B & =\, \frac{4\pi}{c}\,\vec J.
\end{align}
Finally, the electric field is obtained from the generalized Ohm's law~\citep{Valentini2007}:
\begin{multline}
\big(1-d_e^2\nabla^2\big)\vec E = -\frac{\vec u_e\times\vec B}{c} -\frac{\nabla p_e}{n e} + \eta\vec J \\
-\frac{m_e}{m_i}\left\{\frac{\vec u_i\times\vec B}{c} 
   - \frac{1}{ne}\nabla\cdot\Big[m_in(\vec u_i\vec u_i-\vec u_e\vec u_e)+\boldsymbol{\Pi}_i\Big]\right\},\label{eq:ohm}
\end{multline}
where $d_e$, $m_e/m_i$, $\eta$, $n$, $\vec u_i$, $\vec u_e=\vec u_i - \vec J/en$, 
$p_e$, and $\boldsymbol{\Pi}_i$ are the electron skin depth, electron-ion mass ratio, (numerical) resistivity, plasma density,
ion fluid velocity, electron fluid velocity, electron pressure, and ion pressure tensor, respectively.  
The HK model is a quasi-neutral theory~\citep{Tronci2015}, $n_i\simeq n_e\equiv n$, 
and therefore it deals only with frequencies much smaller than the electron plasma frequency, $\omega\ll\omega_{pe}$.
We also note that the finite electron mass in expression~\eqref{eq:ohm} is in practice often set to zero, 
which consequently neglects electron inertia and results in a much simplified version of Ohm's law:
\begin{equation}
\vec E = -\frac{\vec u_e\times\vec B}{c} -\frac{\nabla p_e}{n e} +\ \eta\vec J.
\end{equation}

The gyrokinetic model \citep{Frieman1982, Brizard2007} orders fluctuating 
quantities according to a small expansion parameter $\epsilon$. 
In particular, the following ordering is assumed for the fluctuating quantities:
\begin{align}
\delta f_s/F_{0,s} \sim \delta B/B_0 \sim q_s\delta\phi/T_{0,s} \sim \epsilon, \label{eq:gk1}
\end{align}
where $F_{0,s}$ is the background particle distribution function, $\delta f_s$ is the perturbed part of the 
distribution, $\delta\phi$ is the perturbed electrostatic potential, $T_{0,s}$ is the background 
kinetic temperature (measured in energy units), and $B_0$ is 
the background magnetic field. A similar ordering is assumed for the characteristic macroscopic length scale $L_0$ and 
the frequencies of the fluctuating quantities $\omega$:
\begin{align}
\omega/\Omega_{cs} \sim \rho_s/L_0\sim\epsilon, \label{eq:gk2}
\end{align} 
where $\Omega_{cs}$ is the species cyclotron frequency and $\rho_s$ is the species Larmor radius. For the problems of interest,
such as kinetic-scale turbulence in the solar wind \citep{Howes2006, Howes2008, Schekochihin2009}, 
the macroscopic length scale can be typically associated with 
a characteristic wavelength of the fluctuations measured along the magnetic field.
Time scales comparable to the cyclotron period are cancelled out by performing an 
averaging operation over the particle Larmor motion. The phase space is then conveniently described in 
terms of the coordinates $(\vec R_s, \mu_s, v_{\parallel})$, where $\vec R_s=\vec r + (\vec v\times\vec b_0)/\Omega_{cs}$ is the particle gyrocenter position, $\mu_s=m_s v_{\perp}^2/2B_0$ is the 
magnetic moment, and $v_{\parallel}=\vec v\cdot\vec b_0$ the velocity along the mean field, where $\vec b_0 = \vec B_0/B_0$. 
Instead of evolving the perturbed 
distribution $\delta f_{s}$, the gyrokinetic equations are typically written in terms of the ring distribution:
\begin{align}
h_s(\vec R_s,\mu_s,v_{\parallel},t) = \delta f_{s}(\vec r, \vec v,t) + \frac{q_s{\delta \phi(\vec r,t)}}{T_{0,s}}F_{0,s}(v).  
\end{align}
The distribution $h_s$ is independent of the gyrophase angle at fixed $\vec R_s$. The electromagnetic fields are obtained 
self-consistently from the ring distribution 
under the assumption \eqref{eq:gk1} and \eqref{eq:gk2}. For the electrostatic potential, the assumptions
result in a quasineutrality condition, which is used to determine $\delta\phi$. The magnetic field, on the other hand, 
is obtained from a GK version of Ampere's law. 

Finally, we summarize the main features of the so-called Kinetic 
Reduced Electron Heating Model (KREHM) \citep{Zocco2011}. KREHM is a fluid-kinetic model, obtained as a 
rigorous  limit of gyrokinetics for low-beta magnetized plasmas. In particular, the low beta assumption takes the following form:
\begin{align}
\sqrt{\beta_e} \sim \sqrt{m_e/m_i}\ll 1,
\end{align}
where $\beta_e$ is the electron beta ratio. Due to its low beta limit, the addition of 
KREHM to this work clarifies the plasma beta dependence of our results. Instead of solving for the total 
perturbed electron distribution function $\delta f_e$, KREHM is written in terms of
\begin{align}
g_e = \delta f_e - (\delta n/n_0 + 2v_{\parallel}\delta u_{\parallel,e}/v_{th,e}^2)/F_{0,e},
\label{eq:ge}
\end{align}
where $\delta n_e$ is the perturbed density, $\delta u_{\parallel,e}$ the perturbed electron fluid velocity along the 
mean magnetic field, and $v_{th,e} = \sqrt{2T_{0,e}/m_e}$ is the electron thermal velocity of 
the background distribution $F_{0,e}$. 
The modified distribution $g_e$ has vanishing lowest two (fluid) moments, thus highlighting 
the kinetic physics contained in KREHM, leading to small-scale velocity structures in the
perturbed distribution function. As the final set of equations for $g_e$ does 
not explicitly depend on velocities perpendicular to the magnetic field, 
the perpendicular dynamics may be integrated out, leaving only
the parallel velocity coordinate. As such, the model is by far the least 
computationally demanding of the four that we employ.
As shown in \citet{Zocco2011}, 
a convenient representation of $g_e$ in the $v_{\parallel}$ space 
can be given in terms of Hermite polynomials.

\section{Problem description and simulation setup}
\label{sec:desc}

In the following section we provide details regarding the choice of initial condition and plasma parameters.
Further technical details describing the numerical settings used for each type of simulation are given in 
Appendix~\ref{app:numerics}.

The initial condition used for all simulations 
is the so-called Orszag-Tang vortex \citep{Orszag1979}---a widely-used setup for studying 
decaying plasma turbulence \citep{Biskamp1989, Dahlburg1989, Politano1989, Politano1995, Parashar2009, 
Parashar2015, Loureiro2016, Li2016}. 
The initial fluid velocity and magnetic field perturbation are given by
\begin{align}
\vec u_{\perp} &= \delta u\,\bigl( - \sin(2\pi y/L)\,\hat{\vec e}_x +  
\sin(2\pi x/L)\,\hat{\vec e}_y\bigr), \\
\vec B_{\perp} &= \delta B\,\bigl(-\sin(2\pi y/L)\,\hat{\vec e}_x 
+ \sin(4\pi x/L)\,\hat{\vec e}_y\bigr),
\end{align}
where $L$ is the size of the periodic domain ($x,y\in[0,L)$), and $\delta u$ and $\delta B$ are 
the initial fluid and magnetic fluctuation amplitudes, respectively. A uniform magnetic field 
$\vec B_0 = B_0\hat{\vec e}_z$ is imposed in the out-of-plane ($z$) direction. In addition to magnetic field fluctuations, 
we initialize a self-consistent electric current in accordance with Ampere's law:
\begin{equation}
J_z = \frac{c}{4\pi}\frac{2\pi\delta B}{L}\bigl(2\cos(4\pi x/L) 
+ \cos(2\pi y/L)\bigr).
\end{equation}
For the FK and GK simulations, we explicitly initialize the parallel current via a locally shifted Maxwellian 
for the electron species. Similarly, we prescribe the perpendicular fluid velocities for the FK and HK models 
by locally shifting the Maxwellian velocity distributions in $v_x$ and $v_y$. 
For GK and KREHM, the same approach 
is not possible due to the gyrotropy assumption imposed on the particle perpendicular velocities. Instead, 
the in-plane fluid velocities for GK and KREHM are set up with a plasma density perturbation, resulting in self-consistent 
electrostatic field that gives rise to perpendicular fluid motions via the $\vec E\times\vec B$ drift \citep{Numata2010, Loureiro2016}. 
In the large-scale and strong guide field limit, the leading order term for the electric field is
\begin{equation}
\vec E_{\perp}\approx - \frac{1}{c}\vec u_{\perp}\times \vec B_0
\label{eq:e_perp}
\end{equation}
and the perturbation is electrostatic. For better consistency with reduced-kinetic models used here, the $\vec E$ field 
for the FK model is initialized according to Eq.~\eqref{eq:e_perp} with a corresponding (small) electron 
density perturbation to satisfy the Poisson equation for the electrostatic potential. Apart from minor density 
fluctuations that account for the electrostatic field, the initial ion and electron densities, as well as temperatures, are chosen to be uniform. Finally, it is important to mention that we perform the HK simulations with the generalized version of Ohm's law 
given by Eq.~\eqref{eq:ohm} which includes electron inertia, and as such features a collisionless mechanism for 
breaking the magnetic field frozen-in flux constraint.

A list of simulation runs with the corresponding plasma parameters and box sizes is given in Table~\ref{tab:main}. The key dimensionless parameters varied 
between the runs are the ion beta
\begin{align}
\beta_i = 8\pi n_0 T_i/B_0^2
\end{align}
and the initial turbulence fluctuation strength
\begin{align}
\epsilon = \delta B/B_0 = \delta u/v_A,
\end{align}
where $v_A=B_0/\sqrt{4\pi n_0 m_i}$ is the Alfv\' en speed. 
Throughout this article, the species thermal 
velocity is defined as $v_{th,s} = \sqrt{2T_s/m_s}$ and the species Larmor 
radius is given by $\rho_s = v_{th,s}/\Omega_{cs}$, where $\Omega_{cs} = eB/(m_s c)$ is the species cyclotron frequency.
Unless otherwise stated, lengths are normalized to the ion inertial length $d_i = \rho_i/\sqrt{\beta_i}$, velocities to the Alfv\' en speed $v_A$, 
masses to the ion mass $m_i$, magnetic field to $B_0$, the electric field to $v_AB_0/c$, 
and density to the background value $n_0$. We take the (integral scale) \emph{eddy turnover time} $\tau_0$ as the 
basic time unit, which we define here as
{
\begin{align}
\tau_{0} = \frac{L}{2\pi\delta u}.
\end{align}
}
The eddy turnover time, being independent of kinetic quantities, is a robust measure allowing for a 
direct comparison of simulations obtained from different plasma models and for variable fluctuation 
levels and box sizes \citep{Parashar2015}. Further simulation details pertaining to each kinetic model are provided in 
Appendix~\ref{app:numerics}.

\begin{table}[htb!]
\centering
\begin{tabular}{l l l l l l }
\hline\hline
\multicolumn{6}{c}{\bf Main plasma parameters}\\
\hline
Run & $\beta_i$ & $\epsilon$ & $m_i/m_e$ & $T_i/T_e$ & $L/d_i$ \\
\hline
A1 & 0.1 & 0.2 & 100 & 1 & $8\pi$ \\
A2 & 0.1 & 0.1 & 100 & 1 & $8\pi$ \\
B1 & 0.5 & 0.3 & 100 & 1 & $8\pi$ \\
B2 & 0.5 & 0.15 & 100 & 1 & $8\pi$ \\ 
\hline\hline
\end{tabular}
\caption{List of simulation runs with their corresponding plasma parameters: ion beta ($\beta_i$), initial turbulence fluctuation level ($\epsilon$), ion-electron mass ratio ($m_i/m_e$), 
ion-electron temperature ratio ($T_i/T_e$), and box size in units of the ion inertial length ($L/d_i$).\label{tab:main}}
\end{table}

\section{Results}
\label{sec:results}

Below we present a detailed analysis of the simulation runs listed in Sec.~\ref{sec:desc}. 
All diagnostics are implemented 
following equivalent technical details for all models 
and the numerical values from all the simulations are normalized in the same way.
For the root-mean-square values of the electric current and for the spectral analysis of 
the fully kinetic PIC simulations, we use short-time averaged data as is often
done in the PIC method \citep{Liu2013, Daughton2014, Roytershteyn2015, Munoz2017}. 
The effect of short-time averaging is to highlight the turbulent structures 
and reduce the amount of fluctuations arising from PIC noise that are mainly concentrated at high frequencies. 
In our case, the averaging roughly retains frequencies up to $\omega\lesssim 2\pi\Omega_{ci}$ for all simulation 
runs listed in Table~\ref{tab:main}. The possibility to search for phenomena 
that could potentially reach frequencies much higher than the ion cyclotron 
frequency [e.g.~whistler waves~\citep{Saito2008, Gary2009, Chang2014, Gary2016}] 
is therefore limited in the time averaged data. We note however that without the use of 
short-time averaging the amount of 
PIC noise in our simulations is too high to allow for a detailed analysis of turbulent structures 
at electron kinetic length scales. The limitations set by the PIC noise are most 
strict for the perpendicular electric field, the detailed knowledge of which is an important piece of information
for identifying the small-scale nature of the turbulent cascade. Additional information regarding 
the short-time averaging of the PIC simulation data is given in Appendix~\ref{app:noise}.

\subsection{Spatial field structure}
\label{sec:spatial_structure}

\begin{figure*}[htb!]
\epsscale{1.1}
\plottwo{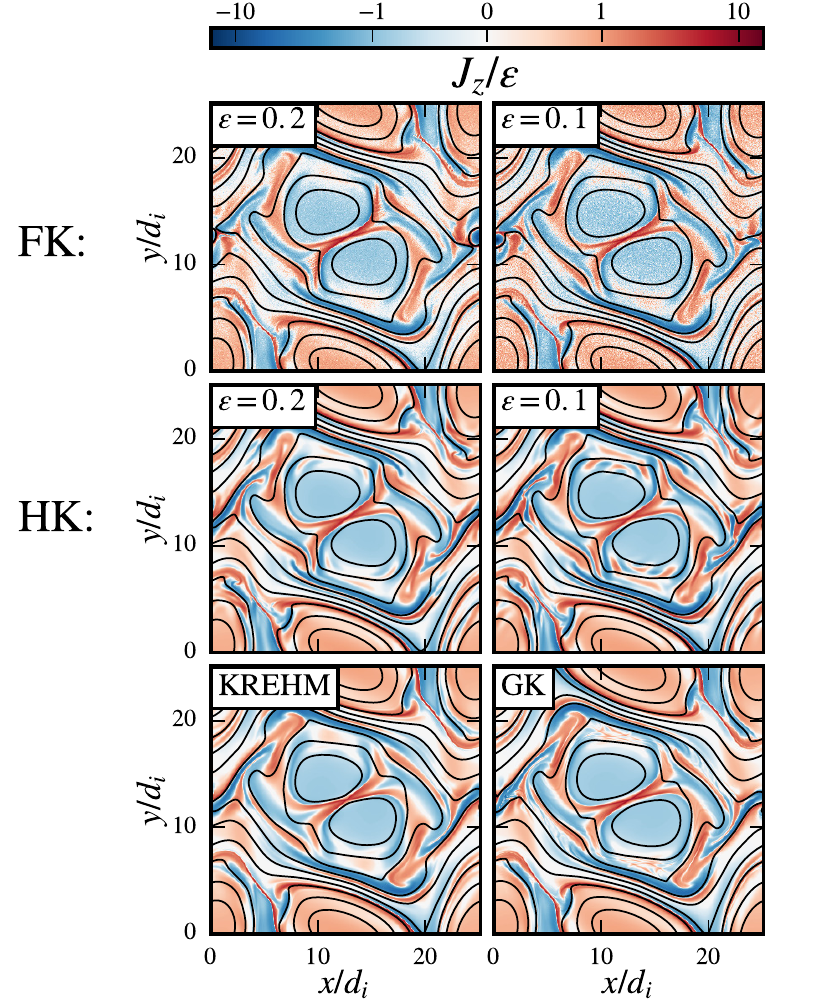}{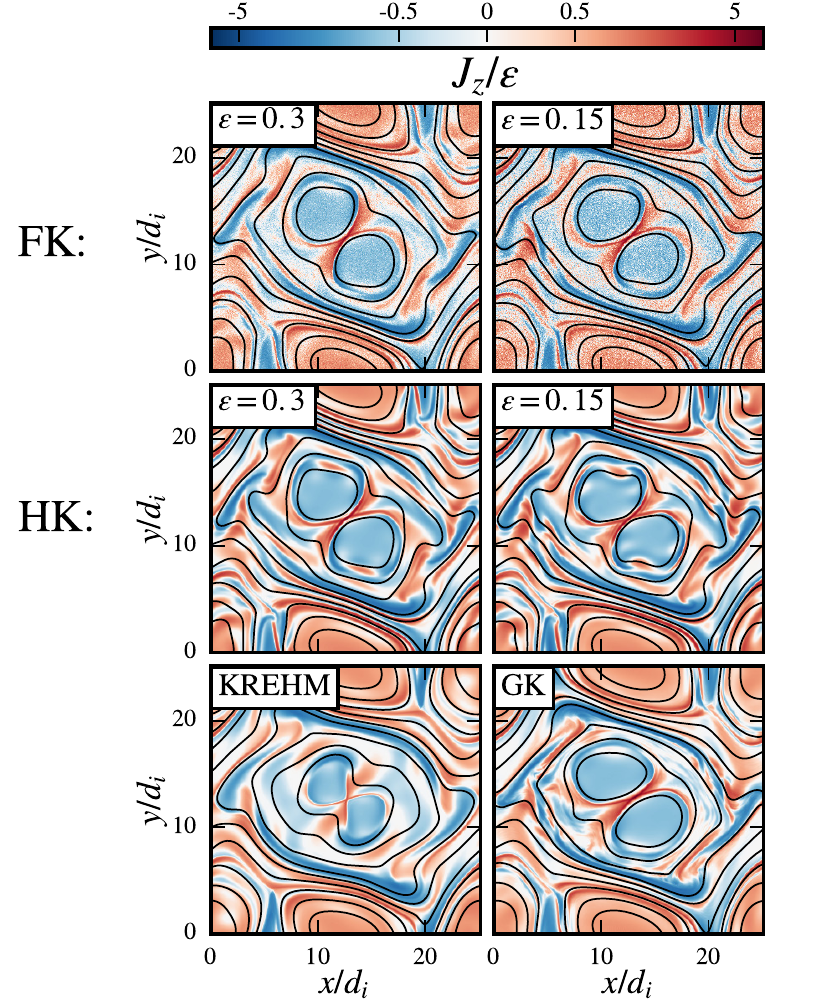}
\caption{Contour plots of $J_z$ for $\beta_i=0.1$ (left) and $\beta_i=0.5$ (right). A 
doubly-logarithmic color scale (one for positive and one for negative 
values, with a narrow linear scale around 0 to connect the two) is used here to reveal 
also the fluctuations of weaker intensity. The color range has been adjusted according to the maximum 
value for any data set.\label{pic:Jz_contours}}
\end{figure*}

We begin the discussion of our results by comparing the spatial structure of the turbulent fields. In Figure~\ref{pic:Jz_contours} we 
compare the out-of-plane electric current $J_z$ at around 3.1 eddy turnover times $\tau_0$ in the simulation. The contour plots 
are shown for variable fluctuation strengths $\epsilon=\delta B/B_0 = \delta u/v_A$ and for both values of the ion beta ($\beta_i=0.1,0.5$).
The electric current obtained from FK simulations is plotted using the raw data without any 
short-time averaging or low-pass filtering to provide the reader with a qualitative measure of the strength of
background thermal fluctuations stemming from PIC noise. To highlight the $\epsilon$ dependence, the FK and HK data
have been rescaled by $1/\epsilon$, whereas in the GK and KREHM models the current is 
already naturally rescaled by $1/\epsilon$. 
With the exception of the KREHM solution corresponding to the $\beta_i=0.5$ case,  
a relatively good overall agreement is found between all models. Since KREHM relies on the low plasma
beta assumption, the disagreement for $\beta_i = 0.5$ is to be expected and highlights the influence of $\beta_i$ 
on the turbulent field structure. On the other hand, KREHM gives 
surprisingly accurate results already for $\beta_i=0.1$ 
even though the formal requirement for its validity 
is given by $\sqrt{\beta_e}\sim\sqrt{m_e/m_i}\ll 1$ \citep{Zocco2011}, 
which gives $\beta_i \lesssim 0.01$ in our case (using $m_e/m_i = 0.01$
and $T_{0,i} = T_{0,e}$). By looking at Fig.~\ref{pic:Jz_contours} we also see 
that the field structure changes relatively little with $\epsilon$. 
Moreover, the field morphology depends much more on the plasma beta than it does on the turbulence fluctuation level. 
It is worth noticing that for even higher $\epsilon$ the field structure 
is expected to change significantly; at least when the sonic Mach number $M_s\sim \epsilon/\sqrt{\beta_i}$ 
approaches unity \citep{Picone1991}.

\begin{figure}[htb!]
\epsscale{1.08}
\plotone{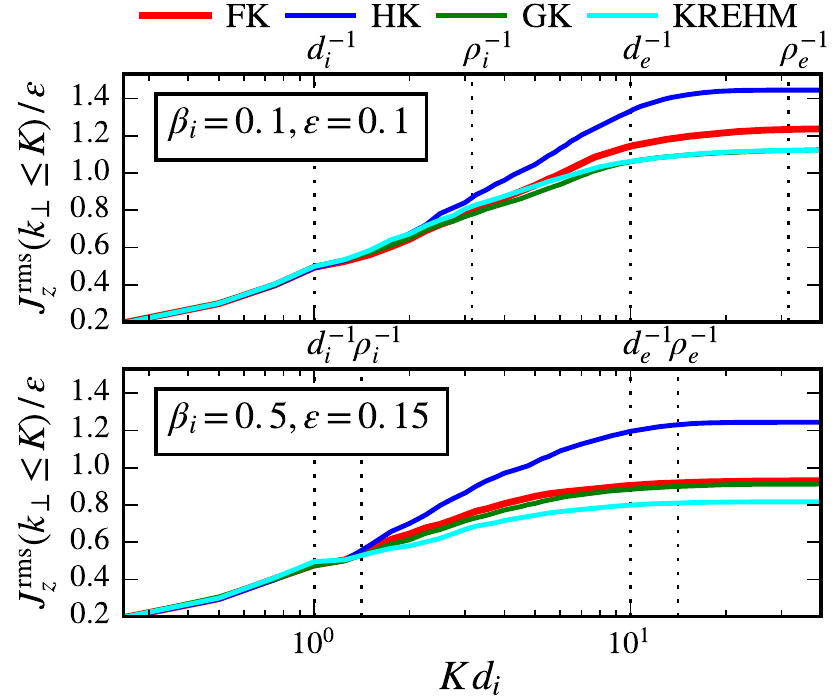}
\caption{\label{pic:j_rms_filt} 
Comparison of the root-mean-square values of the low-pass filtered $J_z$, plotted against the cutoff wavenumber
$K$.}
\end{figure}

In Table~\ref{tab:jz} we list 
the root-mean-square values of the out-of-plane current density at 
4.7 eddy turnover times. Short-time averaged data is used here to compute 
the root-mean-square values for the 
FK model in order to reduce contributions from the background PIC noise. 
While the spatial field structures are overall in good qualitative agreement, the 
root-mean-square current numerical values differ. More specifically, the numerical values 
deviate from the FK model up to 9\% for GK, up to 12\% for KREHM, and up to 33\% for the HK model. We note that 
minor deviations can occur not only as a result of physical differences but also due to differences in the 
numerical grid size and low-pass filters, as well as due to PIC noise in the 
FK runs that may slightly affect the small-scale dynamics. On the other hand, 
regarding the relatively large disagreement between the HK and FK models, it is reasonable
to assume that the main cause for the difference are electron kinetic effects 
absent in the HK approximation. To support the claim,
we consider in Fig.~\ref{pic:j_rms_filt} the root-mean-square values of the low-pass filtered electric current, 
$J_z^{\rm rms}(k_{\perp} \leq K)$, plotted versus the cutoff wavenumber $K$. A significant difference between the 
HK and FK results for $J_z^{\rm rms}$ is generated over the range of scales $1/\rho_i \lesssim k_{\perp} \lesssim 1/d_e$, over which 
the HK turbulent spectra are likely shallower due to the absence of electron Landau damping as demonstrated in Sec.~\ref{sec:spectra}.

\begin{table}[htb!]
\centering
\begin{tabular}{ccccc}
\hline
\hline
\multicolumn{5}{c}{$J_z^{\rm rms}/\epsilon$}\\
\hline
 & \multicolumn{2}{c}{$\beta_i=0.1$} & \multicolumn{2}{c}{$\beta_i=0.5$} \\
 & $\epsilon=0.2$ & $\epsilon=0.1$ & $\epsilon=0.3$  & $\epsilon=0.15$ \\
\hline
FK & 1.09 & 1.24 & 0.85 & 0.93 \\
HK & 1.31 & 1.44 & 1.08 & 1.24 \\
GK & \multicolumn{2}{c}{1.13} & \multicolumn{2}{c}{0.91} \\
KREHM & \multicolumn{2}{c}{1.12} & \multicolumn{2}{c}{0.82} \\
\hline\hline
\end{tabular}
\caption{Comparison of the root-mean-square values of $J_z/\epsilon$ around 4.7 eddy turnover times in the simulation.\label{tab:jz}}
\end{table}

\begin{figure}[htb!]
\epsscale{1.08}
\plotone{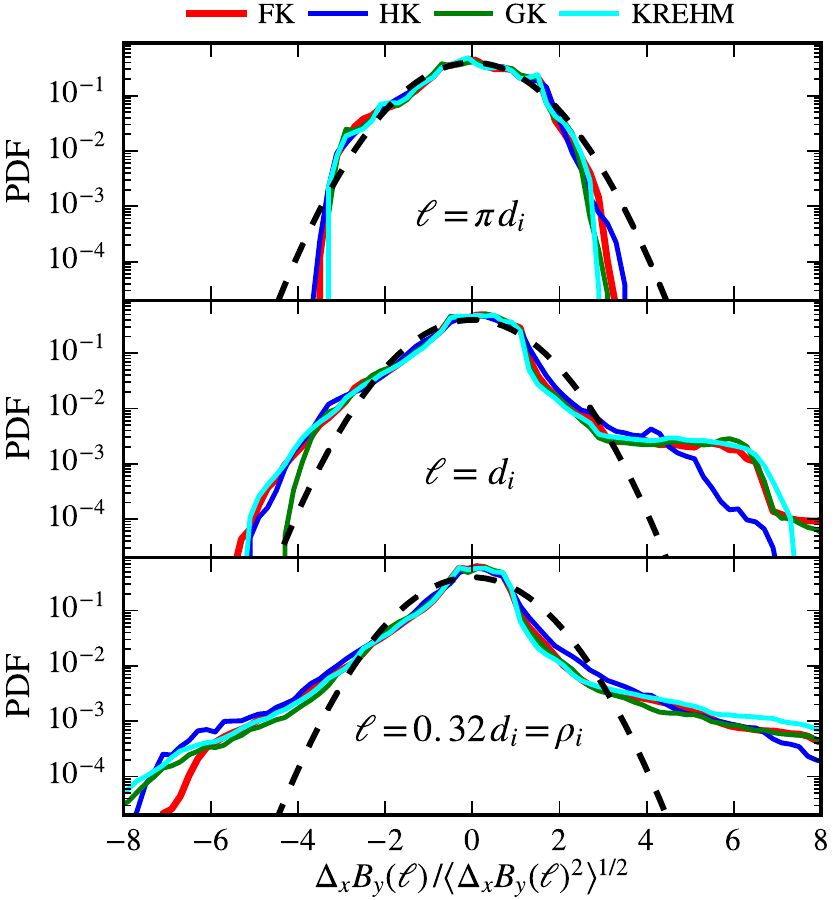}
\caption{PDFs of magnetic field increments for different separation lengths $\ell$. The data is 
shown for the $\beta_i=0.1$ case with $\epsilon=0.1$ for the FK and HK models. The black dashed line corresponds to the 
(normalized) Gaussian PDF.\label{pic:B_incr}}
\end{figure}

In agreement with previous works (see review by~\citet{Matthaeus2015} and references therein), 
the snapshots of $J_z$ in Fig.~\ref{pic:Jz_contours} also reveal a hierarchy of 
current sheets which differentiate with respect to the background Gaussian fluctuations, leading to scale-dependent
statistics also known as intermittency. To investigate intermittency in our simulations, we study the statistics of
magnetic field increments 
$\Delta_x B_{y}({{\ell}}) =  B_{y}(\vec r\nobreak+\nobreak{{\ell}\,\hat{\vec e}_x}) - B_{y}(\vec r)$, using 
the $B_y$ component and with spatial displacements $\ell$ along the (positive) $x$ direction. 
In Figure~\ref{pic:B_incr}, we compare the probability distribution functions (PDFs) of magnetic field increments for the 
$\beta_i=0.1$ runs. 
The PDFs are computed as normalized histograms of $\Delta_x B_{y}({{\ell}})$ with equally spaced bins,
using raw simulation data for the FK model. We also confirmed that 
the PDFs of $\Delta_x B_{y}({{\ell}})$ change only very little upon replacing the raw FK PIC data with 
short-time averaged data (not shown here). To smooth out the statistical fluctuations 
in the PDFs during the turbulent decay, we average the 
PDFs over a time window from roughly 4.4 to 5 eddy turnover times, when 
the turbulence is already well developed.
The departure from a Gaussian PDF (black dashed lines in Fig.~\ref{pic:B_incr}) 
increases from large to small scales. Good agreement is found between all models, except for 
very large values at the tails of the PDFs. Here we point out that the tails 
of the PDFs are likely to be affected by the differences in the smallest resolved scale in each model, which 
depends on the numerical resolution.
Even though the level of intermittency in our simulations, 
compared to the solar wind, is likely underestimated due to a limited system size, the analysis shows 
that the turbulent fields still maintain a certain degree of intermittency, which does not appear to be particularly 
sensitive to different types of reduced-kinetic approximations. Our results are also in qualitative 
agreement with some observational studies and kinetic simulations \citep{Leonardis2013, Wu2013a, Franci2015, Leonardis2016}, 
even though it should be mentioned that some contradictory results have been reported based 
on observational data \citep{Kiyani2009, Chen2014}, for which the authors 
found non-Gaussian but scale-independent PDFs at kinetic scales.

\begin{figure}[htb!]
\epsscale{1.08}
\plotone{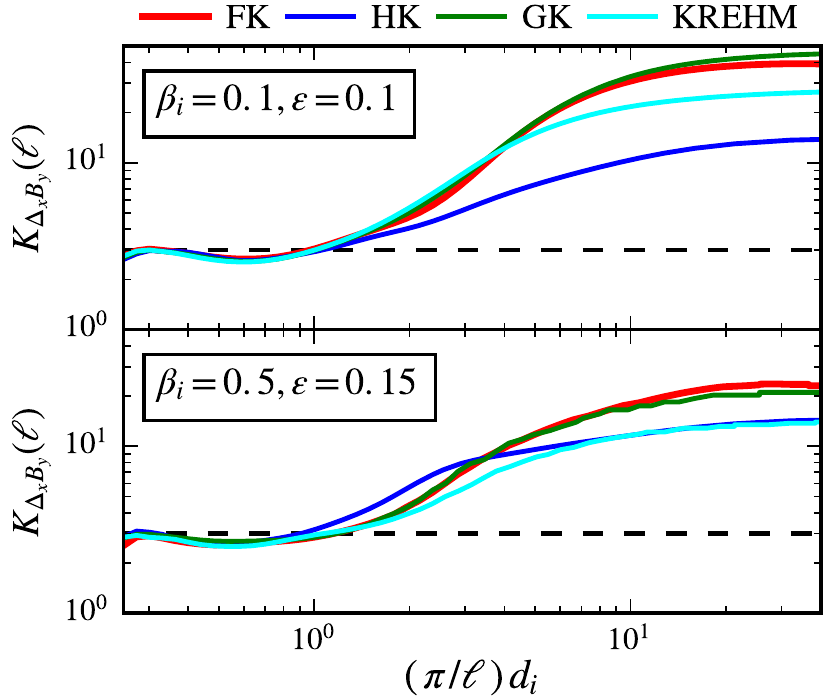}
\caption{Scale-dependent flatness of perpendicular magnetic field increments for $\beta_i=0.1$ (top) and 
$\beta_i=0.5$ (bottom). \label{pic:B_kurt}}
\end{figure}

To compare the level of intermittency in detail, 
we also compute the scale-dependent 
flatness, $K(\ell) = \langle\Delta_x B_{y}(\ell)^4\rangle/\langle\Delta_x B_{y}(\ell)^2\rangle^2$,
of magnetic field fluctuations (Fig.~\ref{pic:B_kurt}), where $\langle\dots\rangle$ represents a space 
average. Similar to the PDFs, we additionally 
average $K(\ell)$ over a time window from 4.4 to 5 eddy turnover times. The flatness is close to the Gaussian 
value of 3 (black dashed lines in Fig.~\ref{pic:B_kurt}) at large scales and increases well above the 
$K=3$ threshold at kinetic scales. Furthermore, Fig.~\ref{pic:B_kurt} also shows that the small-scale intermittency is 
somewhat underestimated in the HK and KREHM simulations. 
As already pointed out, intermittent properties can be very sensitive to the choice of numerical 
resolution, which might be responsible for the lower level of intermittency in the HK simulations.
In particular, the HK flatness is closer to the FK and GK ones for the $\beta_i=0.5$ run, for which the smallest relevant
scale of the FK and GK models ($\rho_e$) is closer to the limits of the HK resolution than in the $\beta_i=0.1$ case. 
Further investigations will be necessary to establish a thorough understanding of 
kinetic-scale intermittency with respect to different types of reduced-kinetic approximations. 
The need to perform a detailed study of intermittency with emphasis 
on kinetic model comparison has already been recognized by several authors and 
a detailed plan to fulfil this (among others) ambitious goal has been 
recently outlined in the so-called ``Turbulent dissipation challenge'' \citep{Parashar2015a}. 

\subsection{Global time evolution}
\label{sec:energy}

Next, we consider the global turbulence energy budget of the decaying Orszag-Tang vortex. 
For this purpose, we split the bulk energy into 
magnetic $M=\langle B^2/8\pi\rangle$, kinetic ion $K_i=\langle nm_i u_i^2/2\rangle$, 
kinetic electron $K_e=\langle nm_e u_e^2/2\rangle$, 
internal ion $I_i=\langle 3nT_i/2\rangle$, and internal electron $I_e=\langle 3nT_e/2\rangle$
energy density, where $\langle\dots\rangle$ denotes a space average. 
To measure the energy fluctuations occurring as a result of turbulent 
interactions, we consider only the relative changes for each energy channel, 
normalized to the sum of the initial magnetic and kinetic ion energy density, $E_0 = M(t=0) + K_i(t=0)$. 
The time traces for the FK model are obtained by space-averaging the raw simulation 
data without the use of short-time averaged fields. For the GK model and KREHM, we 
define the perturbed internal energy similar to \citet{Li2016} as
\begin{align}
\delta I_{s} = -T_{0,s}\delta S_s - \delta K_s + \int_0^t{\mathcal D_s}\dd t',
\end{align}
where $\delta S_s = -\frac{1}{V}\int\dd^2\vec r\dd^3\vec v\delta f_s^2/2F_{0,s}$ is the perturbed species entropy and 
${\mathcal D}_s$ is the mean species dissipation rate, occurring as a result of (hyper)collisional and 
(hyper)diffusive terms in the equations. 
The first two terms on the right-hand side correspond to free energy fluctuations without
contributions from bulk fluid motions. The last term represents the amount of dissipated free 
energy of species $s$. In full $f$ models, the ``dissipated'' nonthermal energy stemming from (numerical) collisional effects 
in velocity space remains part of the distribution function in the form of thermal fluctuations. On the other hand,
in $\delta f$ models such as GK and KREHM the thermalized energy is not accounted for in the background distribution
$F_{0,s}$ and has to be added explicitly to the internal energy expression. The dissipative term is significant
for both ions and electrons. In the GK simulations, hypercollisionality 
contributes to around 56\% and 67\% of $\delta I_e$ and to about 5\% and 
14\% of $\delta I_i$ at $t\approx 6.3\tau_0$ for $\beta_i=0.1$ and $\beta_i=0.5$, respectively. 

\begin{figure*}[hbt!]
\epsscale{0.98}
\plottwo{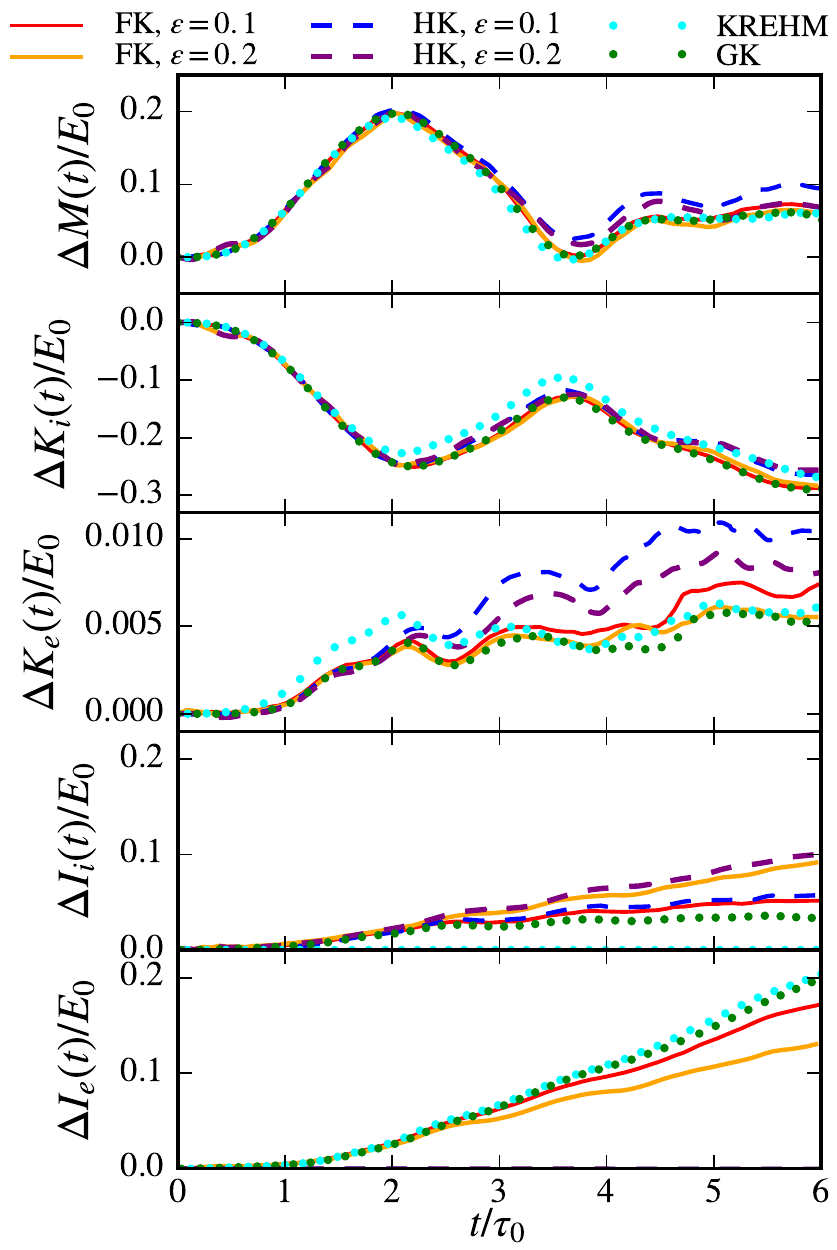}{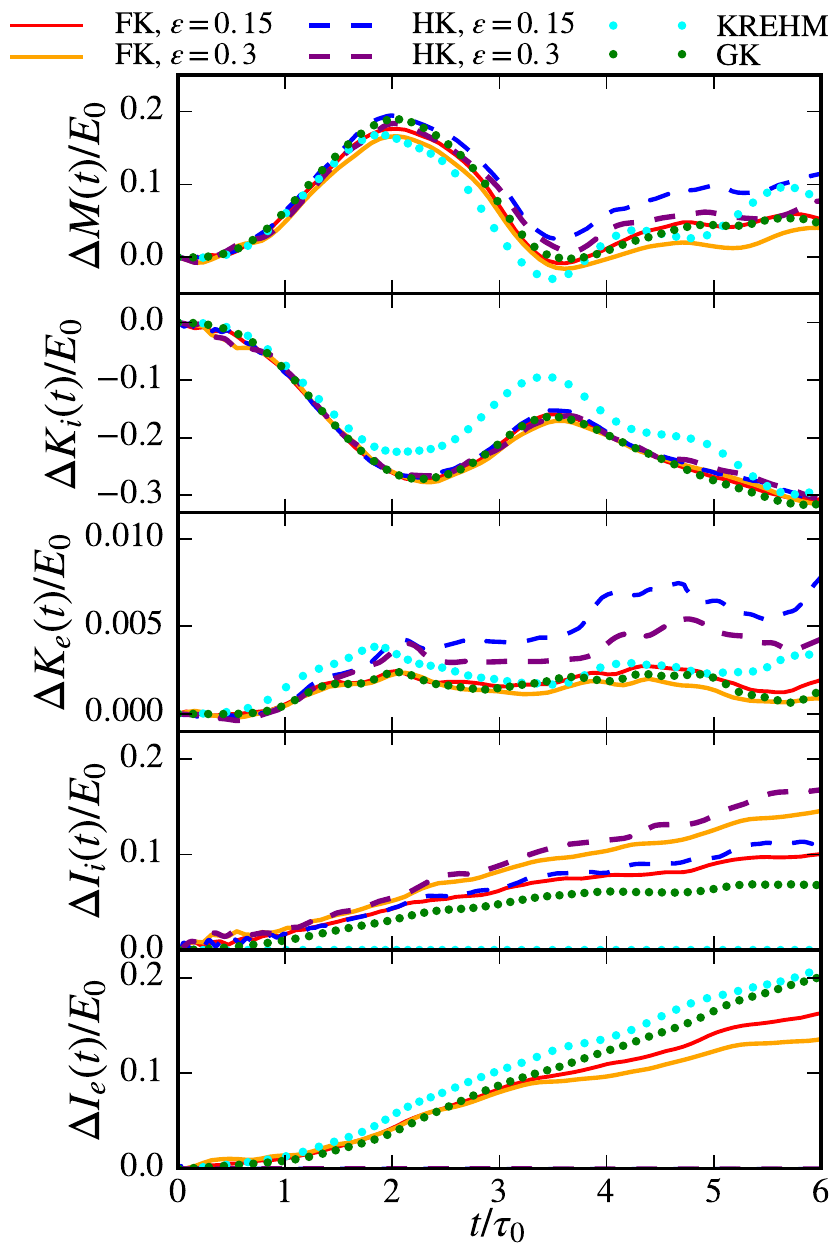}
\caption{Global energy time traces for $\beta_i=0.1$ (left) and $\beta_i=0.5$ (right). Shown from top to bottom are 
the relative changes in the magnetic, kinetic ion, kinetic electron, internal ion, and internal electron energy densities.
All quantities are normalized to the initial turbulent energy $E_0$, which we define as 
the sum of magnetic and ion kinetic energy density.\label{pic:energy_traces}}
\end{figure*}

The time traces of the mean energy densities are shown in Fig.~\ref{pic:energy_traces}. 
The observed deviations from the FK curves appear to be consistent with the 
kinetic approximations made in each of the other three models. In particular,
the HK model exhibits an excess of magnetic and electron kinetic energy density, which can be understood in the 
context of the isothermal electron closure that prevents turbulent energy dissipation via the electron channel.
For the GK model, the result shows that GK tends to overestimate the electron internal energy and underestimate the 
ion one, with the deviations from the FK model growing as $\epsilon$ is increased. 
The observed amplification of ion internal energy relative to the electron one when $\epsilon$ is 
increased is also in good agreement with previous FK studies \citep{Wu2013, Matthaeus2016, Gary2016}.
The convergence of the FK internal energies towards the GK result as $\epsilon\to 0$ is consistent with 
the low $\epsilon$ assumption made in the derivation of the GK equations. Similarly, the improved agreement of KREHM
with the FK model for the lower beta ($\beta_i=0.1$) runs 
can be understood in the context of the low beta assumption made in KREHM. 
We also note that ion heating is ordered out of the KREHM equations, which appears to be reasonable given the 
fact that the ion internal energy increases more rapidly in the higher beta ($\beta_i=0.5$) run. 

In summary, from the point of 
view of the global energy budget, all models deliver accurate results within the formal limits of their validity and in some 
cases even well beyond. The HK model resolves well ion kinetic effects but falls short in 
describing electron features, whereas the GK model becomes increasingly inaccurate for large turbulence fluctuation strengths. 
In addition to the $\epsilon\ll 1$ assumption of GK, the accuracy of KREHM also depends on the smallness of the plasma beta. For the 
particular setup considered, KREHM gives surprisingly good results already for $\beta_i=0.1$, which is well beyond its formal
limit of validity ($\beta_i \lesssim 0.01$ for $m_e/m_i=0.01$ and $T_{0,i} = T_{0,e}$).

\subsection{Spectral properties}
\label{sec:spectra}

The nature of the kinetic-scale, turbulent cascade is investigated by considering the 
one-dimensional (1D) turbulent spectra, which we compute as follows. We divide the 
two-dimensional perpendicular wavenumber plane $(k_x, k_y)$ into equally-spaced
shells of width $\Delta k = k_{\min} = 2\pi/L$ and calculate the 1D spectra, 
$P(k_{\perp})$, by summing the squared amplitudes of the Fourier modes contained in each shell.
The wavenumber coordinates for the 1D spectra are assigned to the middle of 
each shell at integer values of $\Delta k$. To compensate for the global energy variations between different 
models (see Fig.~\ref{pic:energy_traces}) and to account for the fact that the turbulence is decaying, 
we normalize the spectra for each simulation and for each 
time separately, such that $\sum_{k_{\perp}}P(k_{\perp}) = 1$. Finally, we average 
the (normalized) spectral curves over a time interval from 4.4 to 5 eddy turnover times
in order to smooth out the statistical fluctuations in $P(k_{\perp})$.

\begin{figure*}[htb!]
\epsscale{1.}
\plottwo{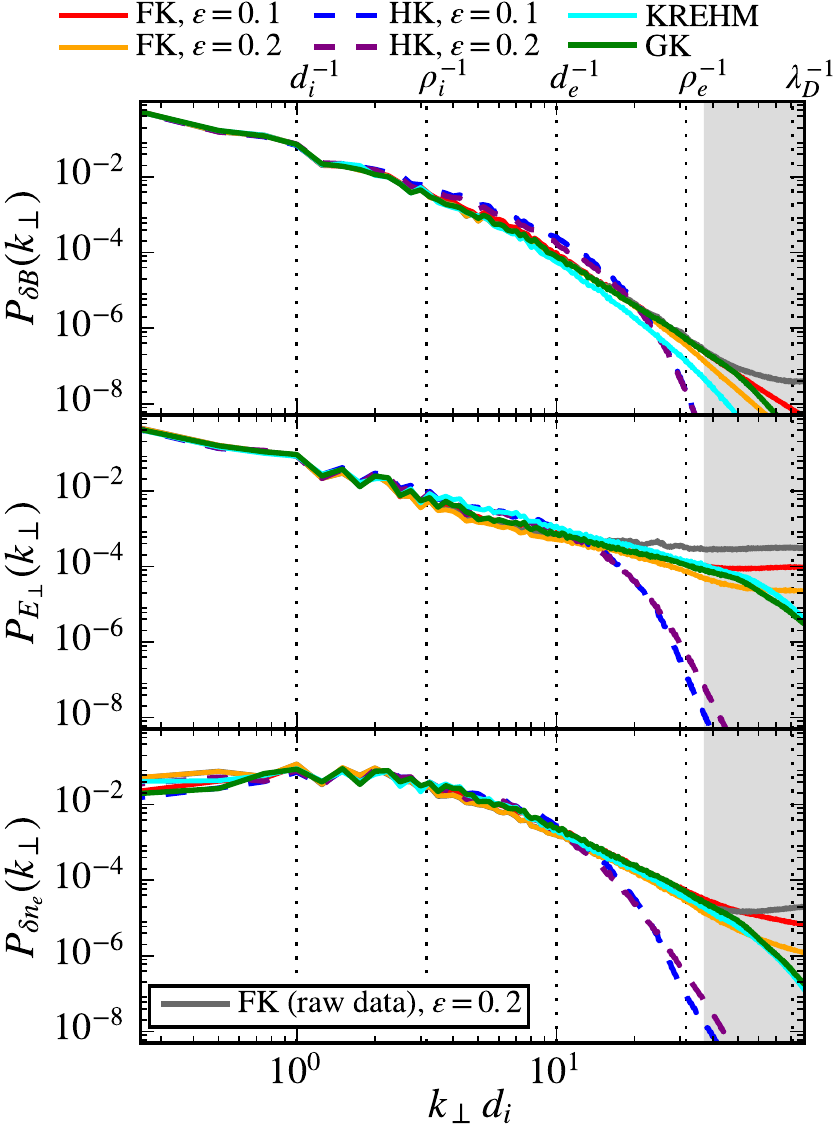}{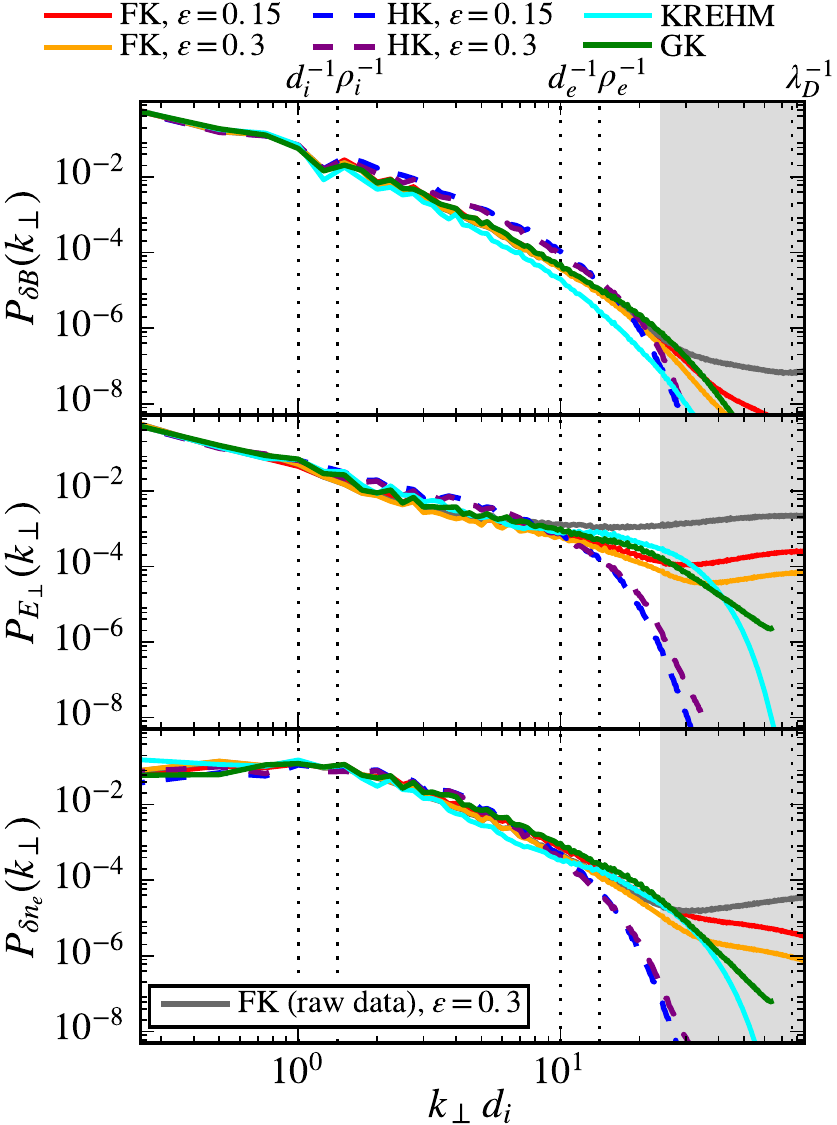}
\caption{One-dimensional $k_{\perp}$ energy spectra for $\beta_i=0.1$ (left) and 
$\beta_i=0.5$ (right). Shown from top to bottom are the $\delta\vec B$ spectra, 
the $\vec E_{\perp}$ spectra, and the electron density spectra. The grey-shaded regions indicate 
the range where PIC noise dominates over the turbulent fluctuations 
for the short-time averaged fully kinetic PIC data. The grey lines 
show the FK spectra for the raw data. \label{pic:spectra}}
\end{figure*}

The turbulent spectra of the magnetic, perpendicular electric, and electron 
density fields are compared in Fig.~\ref{pic:spectra}. The spectra are shown
for all simulation runs, using short-time averaged data for the FK model. For 
reference, we also plot the FK spectra obtained from the raw PIC data in the 
higher $\epsilon$ runs (A1 and B1 in Table~\ref{tab:main}).\footnote{The raw PIC $E_{\perp}$ spectra have been rescaled by 
$\overline{\langle E_{\perp,\rm raw}^2\rangle}/\overline{\langle E_{\perp,\rm avg}^2\rangle}$,
where $E_{\perp,\rm raw}$ is the raw field, $E_{\perp,\rm avg}$ is the short-time averaged field,
and the overline represents a time average from 4.4 to 5 eddy turnover times. The rescaling is used
to compensate for the large amount of PIC noise, which significantly 
affects the average amount of energy in the perpendicular electric field.} 
Since different 
numerical resolutions were used for different models, the scales at which (artificial)
numerical effects become dominant differ between the models. For the HK
model, it can be inferred from the spectrum that numerical effects due to 
low-pass filters~\citep{Lele1992} become dominant for $k_{\perp} \gtrsim 1/d_e$.
On the other hand, for the FK, GK model, and KREHM, the physically 
well-resolved scales are roughly limited to $k_{\perp}\lesssim 1/\rho_e$. In the GK
model and KREHM, the collisionless cascade is terminated by hyperdiffusive terms. In the
FK runs, the main limiting factor are the background thermal fluctuations, which 
dominate over the collisionless turbulence at high $k_{\perp}$, as indicated by the 
grey shading in Fig.~\ref{pic:spectra}.

Two notable features stand out when comparing the spectra over the 
range of well-resolved scales. First, good agreement 
is seen between the FK and GK spectra, and secondly, 
the HK spectra have shallower spectral slopes at {sub-}ion scales. Under the assumptions 
made in the derivation of GK, 
the kinetic-scale, electromagnetic cascade can be understood 
in the context of kinetic Alfv\' en wave (KAW)
turbulence---the nonlinear interaction among quasi-perpendicularly propagating 
KAW packets~\citep{Howes2008, Schekochihin2009, Boldyrev2013}. Thus, the comparison between the 
GK and FK turbulent spectra suggests a predominantly KAW type of turbulent cascade 
between ion and electron scales without major modifications due to physics not 
included in GK, such as ion cyclotron resonance \citep{Li2001, Markovskii2006, He2015}, high frequency waves 
\citep{Gary2004, Gary2009, Verscharen2012, Podesta2012}, and large 
fluctuations in the ion and electron distribution functions. We emphasize here that our results are, 
strictly speaking, valid only for the 
particular setup considered and could be modified in three-dimensional geometry and/or for a 
realistic value of the ion-electron mass ratio. Concerning the HK results, the shallower slopes can be regarded as 
manifest for the influence of electron kinetic physics 
on the sub-ion-scale turbulent spectra. In particular, several previous works demonstrated that 
electron Landau damping could be significant even at ion scales of the solar wind
\citep{Howes2008, TenBarge2013, Told2015, Told2016, BanonNavarro2016}, and 
therefore, we explore this aspect further in what follows.

\begin{figure}[htb!]
\epsscale{1.}
\plotone{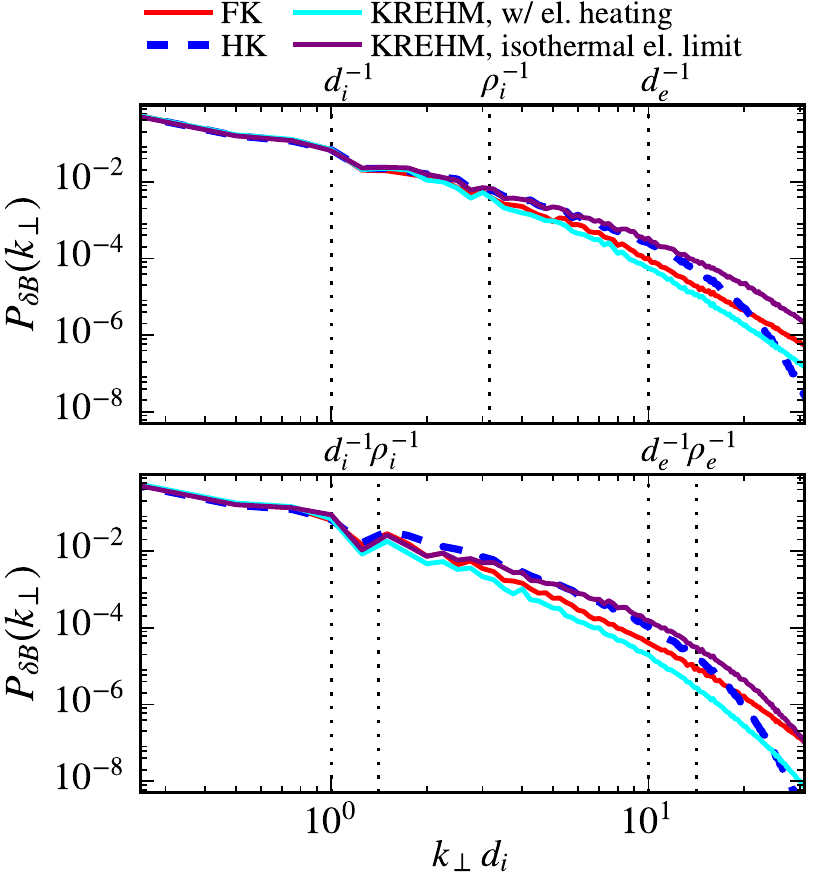}
\caption{Comparison of the FK, HK, and KREHM magnetic field spectra with the isothermal electron limit of 
KREHM for $\beta_i=0.1$ (top) and $\beta_i=0.5$ (bottom).
\label{pic:spec_isotherm}}
\end{figure}

\begin{figure}[htb!]
\epsscale{1.}
\plotone{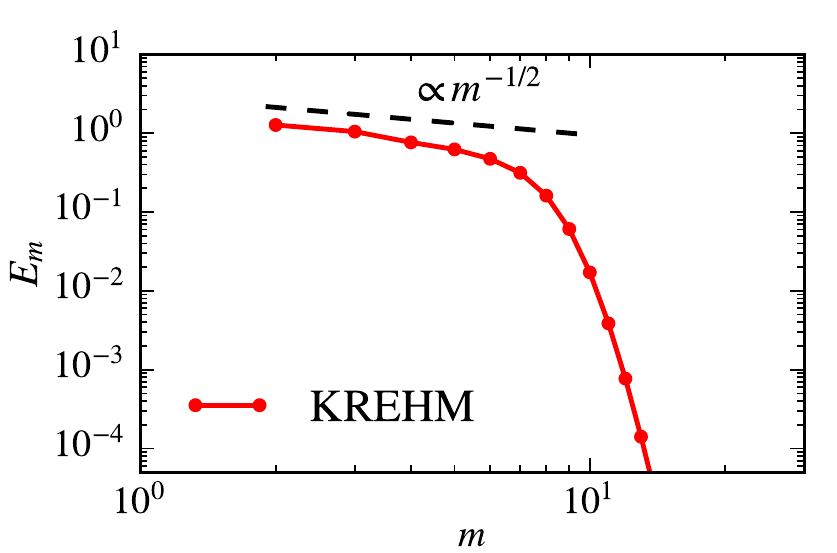}
\caption{Hermite spectrum of the electron free energy, obtained from 
the KREHM simulation corresponding to the $\beta_i=0.1$ set of runs. The $m^{-1/2}$ slope is shown
for reference. \label{pic:spec_m}}
\end{figure}

The role of electron Landau damping is investigated with KREHM, for which electron Landau damping is 
the only process leading to irreversible heating.\footnote{The use of the term irreversible implies here the 
production of entropy, which is ultimately achieved by collisions. For weakly collisional plasmas, such irreversible
heating can only become significant if there exists a kinetic mechanism, such as Landau damping, able 
of generating progressively smaller velocity scales in the perturbed 
distribution function \citep{Howes2008, Schekochihin2009, Loureiro2013, Numata2015, Pezzi2016, BanonNavarro2016}.} 
All other dissipation channels are ordered out as a result of the 
low $\beta$ and low $\epsilon$ limit \citep{Zocco2011}. To test the influence of electron heating on the KREHM
spectra, we performed a new set of simulations in the isothermal electron limit with $g_e$ in Eq.~\eqref{eq:ge}
set equal to $g_e = 0$. In Figure~\ref{pic:spec_isotherm} we compare the FK and HK magnetic field spectra with the 
KREHM spectra obtained from simulations with and without electron heating. Only the curves corresponding to the 
low $\epsilon$ runs (A2 and B2) are shown for the FK and HK models. Over the range of scales where numerical dissipation
in the HK runs is negligible ($k_{\perp}\lesssim 1/d_e$), excellent agreement between the isothermal KREHM spectra 
and the HK spectra is found for the $\beta_i=0.1$ case. On the other hand, when the isothermal electron assumption is
relaxed, the KREHM result is much closer to the FK than to the HK spectrum. A similar trend can be 
inferred for the $\beta_i=0.5$ regime, albeit with some degradation in the accuracy of KREHM, which 
is to be expected given its low $\beta$ assumption. Furthermore, to demonstrate that electron heating indeed 
leads to small-scale parallel velocity space structures, reminiscent of (linear) Landau damping, 
we compute the Hermite energy spectrum of $g_e$ \citep{Schekochihin2009, Zocco2011, 
Loureiro2013, Hatch2014, Schekochihin2016}, which is shown Fig.~\ref{pic:spec_m}.
The Hermite energy spectrum can be regarded as a decomposition of the free energy among parallel
velocity scales (the higher the Hermite mode number $m$, the smaller the parallel velocity scale). Due to a limited 
number of Hermite polynomials used in the KREHM simulations ($M=30$), no clean spectral slope can be observed. However, 
over a limited range of scales ($m\lesssim 7$), the Hermite spectrum is shallow and suggests a tendency 
towards the asymptotic limit $\sim m^{-1/2}$ predicted for linear phase mixing, i.e.~Landau damping 
\citep{Zocco2011}. We also mention that a recent three-dimensional study of the turbulent decay of the Orszag-Tang vortex 
demonstrated a clear convergence towards the $m^{-1/2}$ limit 
for much larger sizes of the Hermite basis (up to $M=100$), thus indicating a certain 
robustness of the result \citep{Fazendeiro2015}.

Finally, the importance of electron Landau damping 
at {sub-}ion scales might be somewhat overestimated in our study due to the use of a reduced ion-electron mass ratio of 100, 
resulting in a $\sim 4.3$ times smaller scale separation between ion and electron scales (compared to a real hydrogen
plasma). In this work, no definitive answer concerning the role of the reduced mass ratio can be given and further 
investigations with higher mass ratios will be needed to clarify this aspect. We note, however, that 
a recent GK study of KAW turbulence, as well as a comparative study of linear wave physics contained in the FK, HK, and GK
model found that electron Landau damping at ion and sub-ion scales can be relevant even for a realistic 
mass ratio \citep{Told2016, BanonNavarro2016}. We also mention that, in principle, one should not overlook
the fact that electron Landau damping is absent in the HK model not only for KAWs but for all other waves as well, 
such as the fast/whistler, slow, and ion Bernstein modes.

To compare the spectral properties of the solutions in even greater detail, we examine 
the ratios of the 1D $k_{\perp}$ spectra \citep{Gary2009, Boldyrev2013, Salem2012, Chen2013, Franci2015, Cerri2016}.
The ratios are often considered in literature as a diagnostic for distinguishing different types 
of wave-like properties (e.g.~KAWs versus whistler waves). We emphasize that the reference to 
wave physics should not be considered as an attempt of demonstrating the dominance of linear dynamics over the 
nonlinear interactions. Instead, the motivation for considering linear properties should be 
associated with the conjecture of critical balance \citep{Goldreich1995, Howes2008, Cho2009, TenBarge2012, Boldyrev2013}, 
which allows for strong nonlinear interactions while still preserving certain properties of the underlying wave physics. 
It is also worth mentioning that a nonzero parallel wavenumber $k_{\parallel}$ along the magnetic field is required 
for the existence of many types of waves (such as KAWs), relevant for the solar wind. 
Wavenumbers with $|k_z| > 0$ along the direction of the mean field are prohibited in our simulations due to 
the two-dimensional geometry adopted in this work. While $k_z = 0$, the magnetic 
fluctuations $\delta B_{K_{\perp}}^{<}$ above some reference scale $\ell_0\sim 1/K_{\perp}$ may 
act as a local guide field on the smaller scales ($\ell\sim 1/k_{\perp} < \ell_0$), 
giving rise to an effective parallel wavenumber 
$k_{\parallel} \sim \vec k_{\perp}\cdot\delta\vec B_{K_{\perp}}^{<}/B_0$ \citep{Howes2008, Cerri2016, Li2016}. 
In this way, certain wave properties may survive even in two-dimensional geometry, even though 
a fully three-dimensional study would surely provide a greater level of physical realism. 

Four different ratios of the 1D spectra are considered: 
\begin{align*}
 C_{A}& = |E_{\perp}|^2/|\delta B_{\perp}|^2,&  C_p & = \beta_i^2\,|\delta n_e|^2/|\delta B_z|^2, \\
 C_{e}& = |\delta n_e|^2/|\delta B|^2,& C_{\parallel} & = |\delta B_z|^2/|\delta B|^2,
\end{align*}
where  $C_e$ and $C_{\parallel}$ are known as the electron and magnetic compressibility, respectively.
Large scale Alfv{\' e}nic fluctuations have $C_A\sim 1$, whereas $C_p \sim 1$ implies a 
balance between the perpendicular kinetic and magnetic pressures. In the asymptotic limit 
$1/\rho_i \ll k_{\perp} \ll 1/\rho_e$, the ratios for KAW fluctuations are expected to behave 
as $C_{A}\sim (k_{\perp}\rho_i)^2/(4 + 4\beta_i)$,
$C_e \sim 1/(\beta_i + 2\beta_i^2)$, $C_{\parallel}\sim \beta_i/ (1 + 2\beta_i)$, and
$C_p\sim 1$, assuming $k_{\perp}\gg k_{\parallel}$, singly charged ions, and $T_{0,i} = T_{0,e}$ 
\citep{Schekochihin2009, Boldyrev2013}. Thus, the $C_A$ 
ratio is expected to grow with $k_{\perp}$, while the other three ratios are in the first approximation roughly 
independent of $k_{\perp}$ on {sub-}ion scales. In the regime $\beta_i\lesssim 1$ relevant to this work, 
we also expect $C_e \gtrsim C_{\parallel}$.

\begin{figure*}[htb!]
\epsscale{1.}
\plottwo{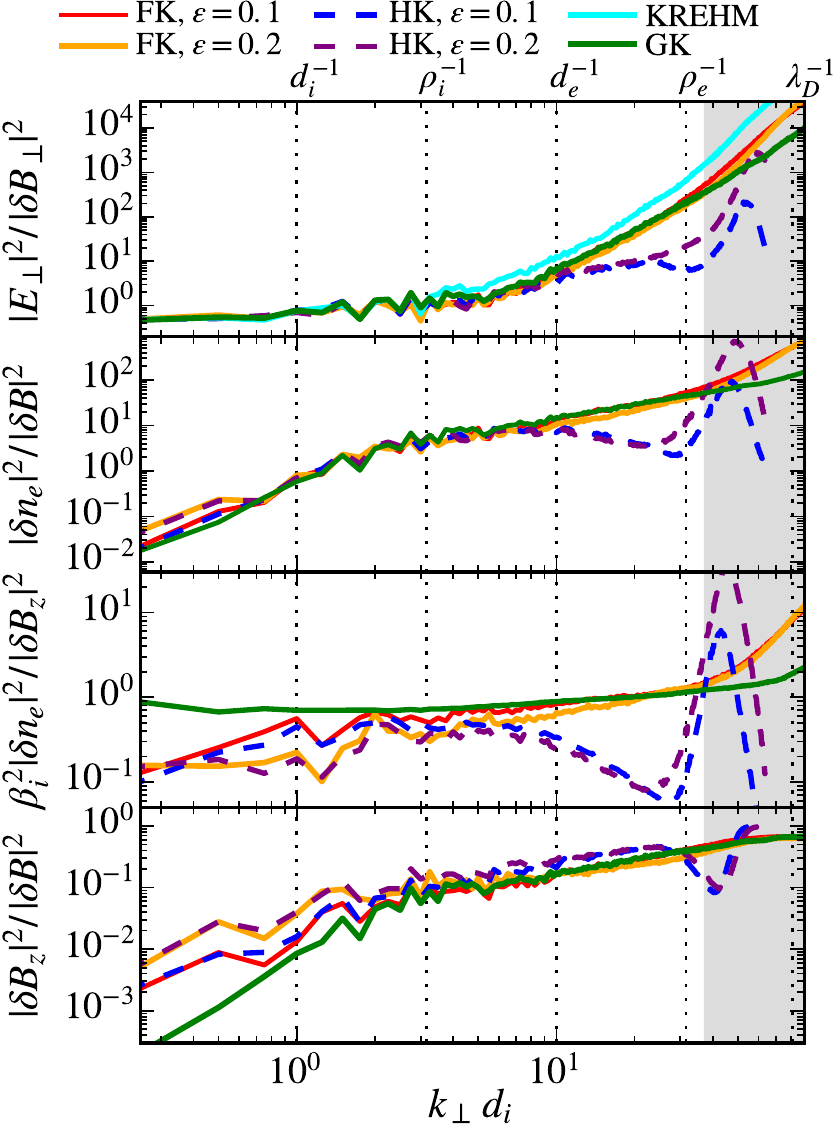}{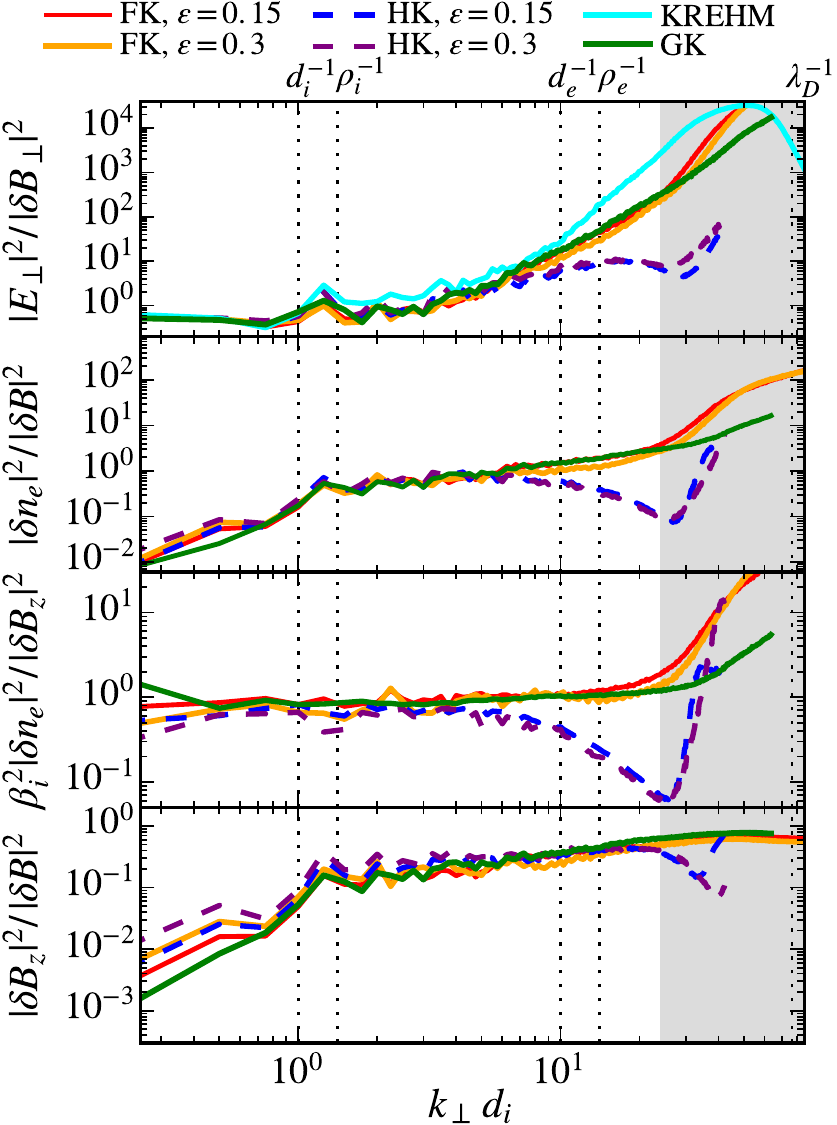}
\caption{Ratios of the one-dimensional turbulent spectra. The left panel shows the ratios 
for the $\beta_i=0.1$ runs and the right 
panel shows the ratios for $\beta_i=0.5$. The grey-shaded regions indicate 
the range of scales dominated by PIC noise in the FK simulations.\label{pic:spectral_ratios}}
\end{figure*}

The spectral ratios are compared in Fig.~\ref{pic:spectral_ratios}. Focusing first on the range 
of scales $1/\rho_i \lesssim k_{\perp} \lesssim 1/\rho_e$, we find good agreement between the GK 
and FK models, albeit with moderate disagreements between the $C_p$ ratios in the $\beta_i=0.1$ regime. 
Thus, the ratios provide firm evidence for a KAW cascade scenario for $\beta_i=0.5$, whereas for 
$\beta_i=0.1$, even though apparently dominated by KAW fluctuations, the cascade is also influenced by 
phenomena excluded in GK. 
Considering the large-scale 
dynamics in the $k_{\perp}\lesssim 1/\rho_i$ range, very good agreement is found between 
the FK and HK results, whereas the GK $C_e$, $C_p$, and $C_{\parallel}$ ratios significantly deviate. 
It is worth noticing that the $C_p$ 
ratio depends only on the magnetic 
fluctuations parallel to the mean field $\delta B_z$, which represent a fraction of the 
total magnetic energy content as evident from the results for $C_{\parallel}$.

The general 
trends exhibited by the FK and HK ratios seem to suggest the presence of high frequency waves 
(e.g. fast magnetosonic modes) excluded in GK. To check if the FK and 
HK solutions in the $k_{\perp}\lesssim 1/\rho_i$ range (and even beyond for $\beta_i=0.1$) 
are indeed consistent with a mixture of 
modes present (KAWs, slow modes, entropy modes) and excluded (e.g.~fast modes) in GK, 
we numerically solve the HK dispersion relation in the $\beta_i = 0.1$ 
regime using a recently developed HK dispersion relation solver \citep{Told2016a}. 
The qualitative findings discussed below for $\beta_i=0.1$ can be also applied to the 
$\beta_i=0.5$ case (not shown here). 

\begin{figure}[htb!]
\epsscale{1.}
\plotone{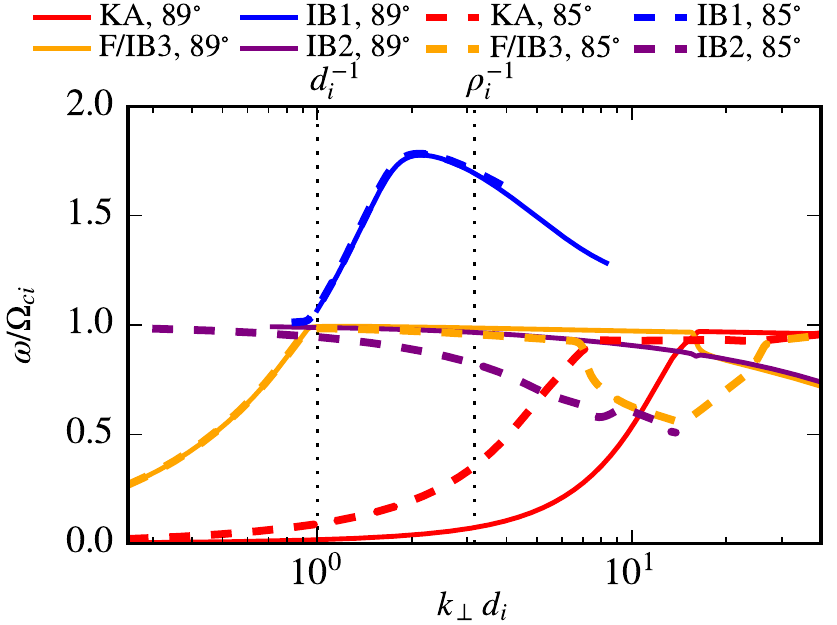}
\caption{\label{pic:disp_rel_omega} Numerical solutions of the hybrid-kinetic dispersion relation for 
$\beta_i=0.1$. The labels KA, F, and IB are used to denote kinetic Alfv{\' e}n  waves, fast waves, 
and (different branches of) ion Bernstein modes, respectively.}
\end{figure}

A collection of least damped modes identified 
with the HK linear dispersion relation solver is 
shown in Fig.~\ref{pic:disp_rel_omega}. Two fixed propagation angles with respect to the magnetic field 
are considered; 89 and 85 degrees. The identified branches can be interpreted as KAWs, fast 
modes, and generalized ion Bernstein modes. For the sake of simplicity, we do not consider  
slow modes and entropy modes because our goal here is to study the properties of 
waves excluded in GK [for a detailed discussion on GK wave physics see \citet{Howes2006}]. 
The identified ion Bernstein modes are generalized in 
the sense that they are not purely perpendicularly propagating nor electrostatic and they also 
split into multiple branches around the cyclotron frequency. As pointed out by 
\citet{Podesta2012}, the fine splitting of the ion Bernstein modes may increase 
the opportunity for wave coupling with the fast and KAW branch. 

\begin{figure}[htb!]
\epsscale{1.}
\plotone{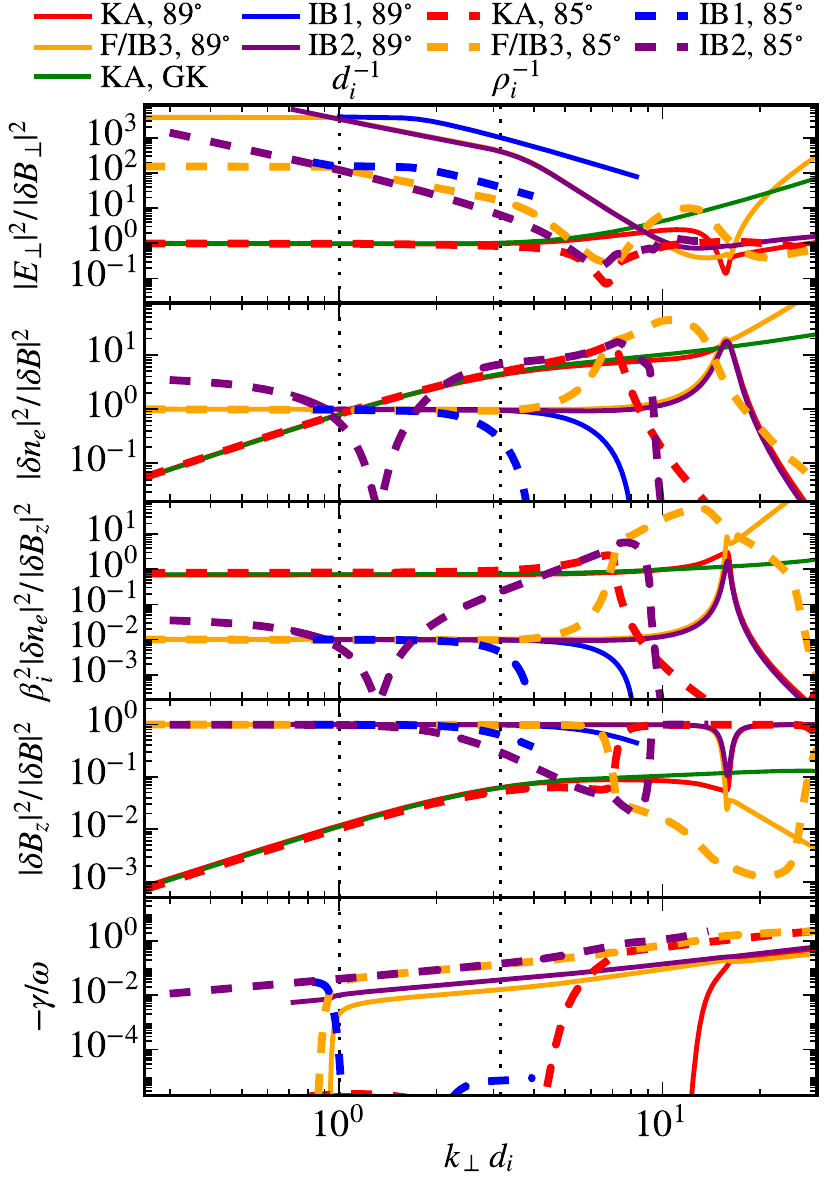}
\caption{\label{pic:ratios_disp_rel} Spectral ratios corresponding to different solutions of the 
HK dispersion relation (top four plots) and their linear damping rates (bottom). 
For reference, we also show the 
GK prediction for the spectral ratios of KAWs.}
\end{figure}

Having identified the main 
branches of interest, we now proceed to show the corresponding spectral ratios (Fig.~\ref{pic:ratios_disp_rel}).
In addition to the spectral ratios obtained from the HK solver, we also show for reference the GK prediction for the 
spectral ratios of KAWs (using $m_i/m_e = 100$), as well as the linear damping rates of the HK branches. 
Comparing the linear predictions with the turbulent ratios shown in the 
left panel of Fig.~\ref{pic:spectral_ratios}, we find that the directions in which the FK and HK turbulent ratios 
are ``pulled away'' from the GK curves are consistent with linear properties of fast and 
ion Bernstein modes. In particular, the magnetic compressibility $C_{\parallel}$ is order unity for the fast 
and ion Bernstein modes, whereas their $C_p$ ratio is below the pressure balance regime 
$C_p\sim 1$. Similarly, the electron compressibility $C_e$ exceeds the one of KAWs at low wavenumbers. All these 
properties are in qualitative agreement with the general trends exhibited by the FK and HK turbulent solutions 
for $k_{\perp}\lesssim 1/\rho_i$. Regarding the $C_A$ ratio, we note that $\delta B_{\perp}/\delta B\ll 1$ 
for the fast and ion Bernstein modes, meaning that these modes cannot significantly 
influence the total $|E_{\perp}|^2/|\delta B_{\perp}|^2$ ratio when mixed 
together with the Alfv{\' e}nic fluctuations. Looking at the damping rates, linear theory predicts cyclotron damping of 
the fast mode already at $k_{\perp} \sim 1/d_i$. However, the GK turbulent ratios deviate 
from the FK model beyond the $k_{\perp} \sim 1/d_i$ scale for $\beta_i=0.1$. 
Therefore, the coupling of the fast modes to the 
ion Bernstein modes is likely significant in the low-beta regime. 
The most natural candidate for the conversion 
of the fast wave would be the mode denoted as IB1 in Fig.~\ref{pic:disp_rel_omega}, which is 
very weakly damped in the HK model and allows for a continuation of the fast branch 
above the ion cyclotron frequency. In the FK model, ion Bernstein waves are subject to additional 
damping due to electron Landau resonance which could potentially explain why all the 
FK ratios show a tendency for converging onto the GK curves with increasing wavenumbers, 
whereas in the HK model, the turbulent ratios (in particular $C_p$ and $C_{\parallel}$) do not share the same trend. 

In summary, we conclude that the deviation of the GK model 
from the FK and HK turbulent solutions in the range $k_{\perp}\lesssim 1/\rho_i$, 
and possibly even beyond for $\beta_i=0.1$, can be reasonably well 
explained with the presence of fast magnetosonic and ion Bernstein modes, coexisting 
together with Alfv{\' e}nic fluctuations. On the other hand, 
we admit that alternative explanations involving, for example, non-wave-like 
phenomena could be in principle possible. As a side note, it is also worth mentioning that 
according to linear theory, KAWs should undergo cyclotron resonance around $k_{\perp}d_i \sim 10$ (see 
bottom plot in Fig.~\ref{pic:ratios_disp_rel}), resulting in an abrupt change in the spectral ratios. 
No such variation is found in the FK turbulent ratios. This suggests that
cyclotron resonance has in our case only a minor effect on the turbulent cascade, in consistency with the 
arguments against cyclotron resonance given by \citet{Howes2008}.

\subsection{Ion and electron nonthermal free energy fluctuations}
\label{sec:free_energy}

At last, motivated by previous works on non-Maxwellian velocity structures in HK simulations 
\citep{Greco2012, Valentini2014, Servidio2015} and free energy cascades in GK and KREHM
\citep{Schekochihin2009, BanonNavarro2011, Zocco2011, Told2015, Schekochihin2016},  
we investigate the spatial distribution of the nonthermal species free energy fluctuations:
\begin{align}
\delta\tilde{\mathcal E}_s = \int\frac{T_{0,s}\delta\tilde f_s^2}{2F_{0,s}}\dd^3\vec v,
\label{eq:free_energy} 
\end{align}
where $\delta \tilde f_s$ is the perturbed part of the distribution function 
with vanishing lowest three moments (density, fluid velocity, and temperature),  
$T_{0,s}$ is the equilibrium temperature, and $F_{0,s}$ is the background Maxwellian.
Expression \eqref{eq:free_energy} can be regarded as a measure for quantifying the 
deviations from local thermodynamic equilibrium, characterized by non-Maxwellian 
fluctuations in velocity space that set the stage for irreversible plasma heating to occur. 
Due to a limited availability of data and refined diagnostic tools, needed for the 
calculation of the nonthermal free energy as a function of space, 
we present here only the results from a 
subset of all simulations and models. In particular, we focus on the $\beta_i=0.1$ regime 
and consider the ion nonthermal fluctuations calculated from the HK model and the 
electron nonthermal fluctuations calculated from the $v_{\perp}$-integrated distribution 
functions in the FK model and KREHM. In the HK and FK models, we define $\delta \tilde f_s$ as 
$\delta \tilde f_s = f_s - F_{0,s}$, where $F_{0,s}$ is the \emph{local} Maxwellian with matching 
density, fluid velocity, and temperatures of the total distribution $f_s$. In KREHM,
the equivalent of expression \eqref{eq:free_energy} can be defined as
$\delta\tilde{\mathcal E}_{\parallel, e} = n_{0}T_{0,e}\sum_{m=3}^M\hat g_m^2/2$,
where $\hat g_m$ are the Hermite expansion coefficients of $g_e$ \citep{Zocco2011, Schekochihin2016}. 
By skipping the lowest three coefficients ($m\leq 2$) in the above sum, the contributions 
from density, fluid velocity, and temperature fluctuations are explicitly 
excluded from the KREHM free energy. Here, we only consider parallel electron velocity fluctuations 
because $g_e$ does not depend on $v_{\perp}$ in KREHM. For the FK model, on the other 
hand, the $v_{\perp}$-integrated electron distribution function is used as a necessity, due to a limitation
in diagnostic tools presently available for the FK PIC code used in this work. The limitation 
to parallel velocity fluctuations is, however, physically well-motivated for the electrons 
based on a number of previous works, which showed that electron heating takes 
place predominantly in the parallel direction \citep{Saito2008, TenBarge2013, Haynes2014, Numata2015, Li2016, BanonNavarro2016}.

\begin{figure*}[htb!]
\epsscale{1.1}
\plotone{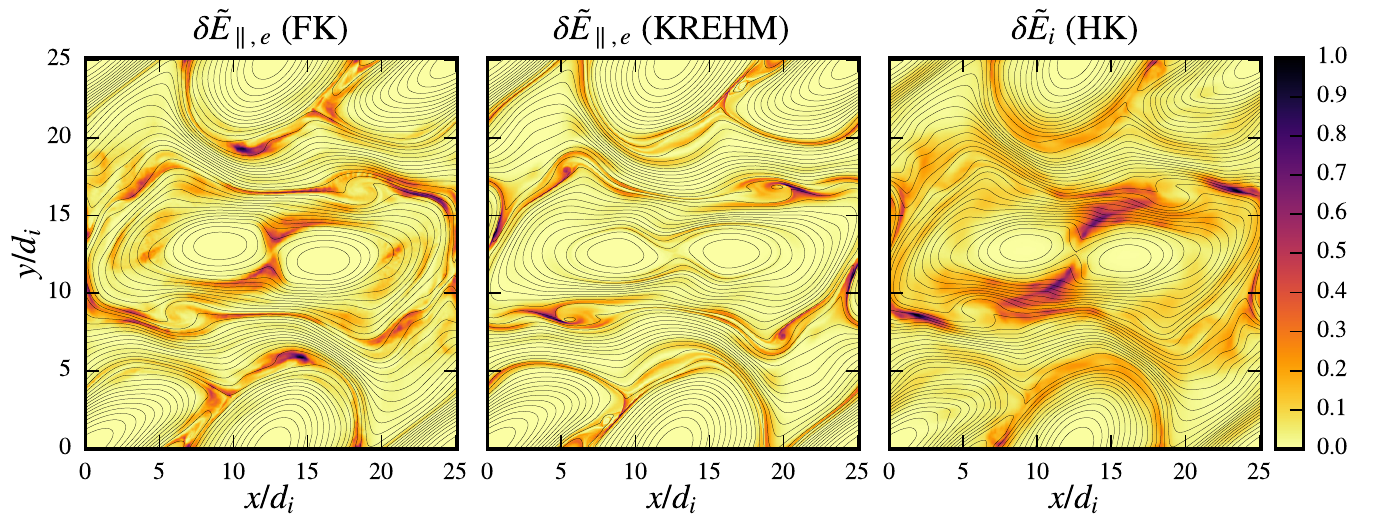}
\caption{Nonthermal electron and ion free energy fluctuations. In the left and middle plots, we compare the 
FK and KREHM electron free energy, stemming from small scale structures in $v_{\parallel}$. 
The plot on the right shows the (total) nonthermal ion free energy 
obtained from the HK simulation. Overplotted are the contours of the vector potential $A_z$. 
The color map is normalized to the maximum value of the free energy, separately for each plot.\label{pic:free_energy}}
\end{figure*}

\begin{figure}[htb!]
\epsscale{1.18}
\plotone{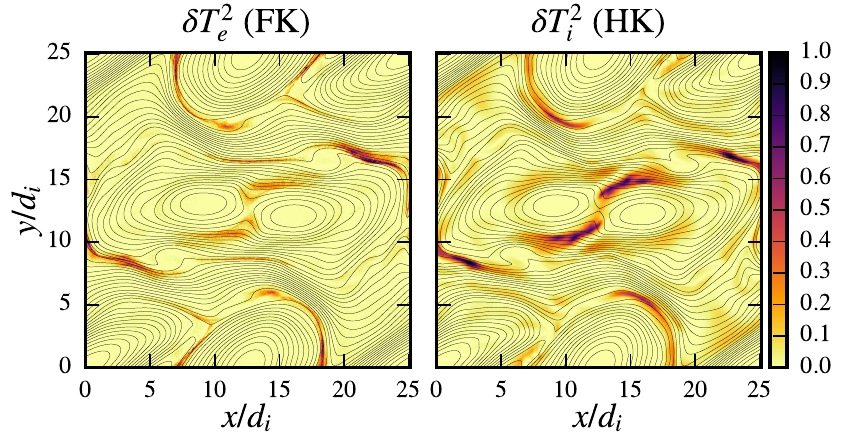}
\caption{Ion and electron temperature fluctuations corresponding to the 
HK and FK nonthermal free energies shown in Fig.~\ref{pic:free_energy}.\label{pic:temp_fluct}}
\end{figure}

The nonthermal ion and electron free energy fluctuations are shown in Fig.~\ref{pic:free_energy}.
The snapshots correspond to the $\beta_i=0.1$ runs around 4.7 eddy turnover times, with $\epsilon=0.2$
in the HK and FK simulations. For reference, we also show in Fig.~\ref{pic:temp_fluct} the 
corresponding ion and electron temperature fluctuations $\delta T_s = T_s - \langle T_s\rangle$, 
plotted at same time in the HK and FK simulation. 
A highly nonuniform spatial distribution of the nonthermal free energy
is found for both ions and electrons. Furthermore, by overplotting the contours of the 
vector potential $A_z$ we find that the nonthermal fluctuations 
are mostly concentrated around small-scale magnetic reconnection sites, corresponding (in 2D) 
to the X-points of $A_z$. Fig.~\ref{pic:free_energy} also 
shows some disagreements in the nonthermal free energy spatial profiles between the FK model and 
KREHM. Considering the abovementioned circumstances, one could suppose 
that the disagreements are related to the small-scale differences in magnetic field configuration, 
which might consequently impact the exact locations where reconnection is taking place. 
A detailed investigation of the possible causes for the observed 
disagreements is beyond the scope of this study but could be carried out in future works. 

The observed correlation between reconnection sites and nonthermal fluctuations 
is in good agreement with previous works 
\citep{Greco2012, Valentini2014, Haynes2014, Servidio2015} and as such reinforces the idea that reconnection
might play a significant role in the turbulent heating of kinetic-scale, collisionless plasma turbulence.
Compared to Fig.~\ref{pic:temp_fluct}, it is seen that 
the nonthermal free energy peaks are well correlated with temperature fluctuations, the main difference to the 
latter being that the nonthermal fluctuations are spatially more diffused. This seems to suggest that 
the energy supplied to the particles at reconnection sites is progressively cascaded to smaller velocity scales 
in the reconnection outflows, thus allowing the nonthermal fluctuations to spread over a larger volume 
compared to the temperature fluctuations. The generation of non-Maxwellian velocity space fluctuations can be, 
on the other hand, linked with the presence of wave-particle interactions 
such as Landau damping, as well as with kinetic instabilities \citep{Haynes2014}.

While the role of Landau damping in kinetic-scale, collisionless plasma turbulence 
is presently somewhat less known for the ions, the significance of electron Landau damping 
has been demonstrated in many previous studies \citep{Howes2008, TenBarge2013, Told2015, BanonNavarro2016} 
and in Sec.~\ref{sec:spectra} of this work. Conversely, we find---at the same time---that 
reconnection could as well be a major process leading to heating in collisionless plasma turbulence. 
Similar as for Landau damping, the latter idea is also well-supported by a number of previous studies 
\citep{Sundkvist2007, Osman2011, Karimabadi2013a, Chasapis2015, Matthaeus2015}. 
Some of the previous works tried to establish a sharp distinction between heating due to 
wave-particle interactions versus heating occurring as a consequence of reconnection, often 
favouring one of the two possibilities as the main cause for turbulent heating.
As one additional option, we would like to promote a rather different view, which is that these two seemingly 
distinct processes might work hand in hand and are spatially entangled with each other, making a 
clear distinction between heating due to Landau damping and reconnection a somewhat ill-posed task. Instead,
it might be more instructive to consider these two processes in a unified framework and investigate, for example, 
how the (nonuniform) damping via Landau resonance might be affected by the presence of reconnection outflows, which
could (locally) modify the Landau resonance condition. This view is supported by previous
studies of weakly-collisional reconnection \citep{Loureiro2013, Numata2015}, and by a recent 
study of GK turbulence \citep{Klein2017} which showed that, contrary to naive expectations, Landau 
damping in a turbulent setting can be spatially highly nonuniform and might be responsible for the intermittent
heating observed in the vicinity of current sheets \citep{Sundkvist2007, Osman2011, Wu2013a, Karimabadi2013a, Chasapis2015}. 
The possibility of a close relationship between reconnection and Landau damping 
was also implied in a recent study by \citet{Parashar2016}, which mentioned a mechanism described 
as a nonlinear generalization of Landau resonance. 

Finally, it is worth acknowledging one 
additional aspect of kinetic-scale reconnection, recently explored by \citet{Cerri2017a}, \citet{Mallet2017}, \citet{Loureiro2017}, and 
\citet{Franci2017}. In the abovementioned works, the authors argue that 
reconnection could also significantly impact the nonlinear energy transfer and the 
shape of the turbulent spectra at kinetic scales. This would consequently imply a more generalized type of kinetic cascade; 
one which is also influenced by non-local energy transfers (in spectral space) as opposed to local mode couplings in wave-like models of 
turbulence \citep{Galtier2003, Howes2008, Schekochihin2009, Boldyrev2013, Passot2015}. A detailed investigation of this novel
cascade phenomenology is beyond the scope of this work. Strictly speaking, the turbulent spectra and spectral ratios alone 
are not sufficient to unambiguously determine whether the energy transfer is local or not. Instead, the spectra and spectral ratios can 
only point out the dominant type of turbulent fluctuations at each wavenumber separately.

\section{Conclusions}
\label{sec:conclusions}

In this work, we performed a detailed comparison of kinetic models in two-dimensional, collisionless plasma turbulence with
emphasis on kinetic-scale dynamics. Four distinct models were included in the comparison: the fully kinetic (FK), 
hybrid-kinetic (HK) with fluid electrons, gyrokinetic (GK), and a reduced gyrokinetic model, formally derived as a low 
beta limit of gyrokinetics (KREHM). Two different ion beta ($\beta_i$) regimes and variable 
turbulence fluctuation amplitudes ($\epsilon$) were considered ($\beta_i=0.1$ using $\epsilon=0.1,0.2$ and 
$\beta_i=0.5$ using $\epsilon=0.15,0.3$). 
The main findings can be summarized as follows:

\begin{itemize}
\item For $\beta_i=0.5$, the kinetic-scale ($k_{\perp}\gtrsim 1/d_i$) 
spectral properties of the FK and GK solutions were found to be in good agreement, 
thus suggesting a kinetic Alfv{\' e}n cascade scenario from ion to electron scales without significant modifications
due to physics not included in GK (Figs.~\ref{pic:spectra} and \ref{pic:spectral_ratios}).
\item A detailed comparison between the GK and FK spectral ratios at $\beta_i=0.1$ reveals a deviation between the 
two models at kinetic scales (Fig.~\ref{pic:spectral_ratios}). However, given that the kinetic-scale ($k_{\perp}\gtrsim 1/d_i$) disagreement 
is clearly observed only for two out of four different spectral ratios considered, nor is it clearly seen in the turbulent 
spectra themselves (Fig.~\ref{pic:spectra}), it is reasonable to assume that kinetic Alfv{\' e}n 
fluctuations still play a significant role even in the low-beta regime.
\item At the largest scales ($k_{\perp}\lesssim 1/d_i$), 
the ratios of the GK turbulent spectra notably deviate from the FK and HK approaches for both ion betas (0.1 and 0.5), 
the likely cause for it being the lack of the fast magnetosonic modes in the GK approximation (Figs.~\ref{pic:spectral_ratios}, \ref{pic:disp_rel_omega}, and \ref{pic:ratios_disp_rel}). 
The disagreement is larger for higher fluctuation amplitudes 
and for the lower value of the ion beta ($\beta_i=0.1$), for which the deviations 
carry over to kinetic scales, possibly via mode coupling to ion Bernstein waves. 
\item The sub-ion-scale HK spectra were found to be shallower than those obtained from the FK and GK simulations, 
presumably due to the lack of electron Landau damping (Figs.~\ref{pic:spectra}, \ref{pic:spec_isotherm}, and \ref{pic:spec_m}).
\item The real space turbulent field structures are in good qualitative agreement (Fig.~\ref{pic:Jz_contours}). 
Furthermore, all models considered give rise to intermittent statistics at kinetic scales, albeit with some 
minor quantitative differences between the models (Figs.~\ref{pic:B_incr} and \ref{pic:B_kurt}). 
The main reason for the observed deviations might be 
related to the fact that different numerical resolutions were used for different models. 
Further studies will be necessary to identify if the observed quantitative differences are indeed physical.
\item The spatial profiles of the nonthermal ion and electron free energy fluctuations suggest that
kinetic-scale reconnection might play an important role in the heating of the plasma 
(Figs.~\ref{pic:free_energy} and \ref{pic:temp_fluct}). Furthermore, our results 
also suggest that reconnection might be closely entangled with (nonuniform) Landau damping, 
making the distinction between heating due to reconnection versus heating due to Landau damping an ill-posed task.
\item KREHM delivers surprisingly accurate results already for $\beta_i=0.1$, 
even though its range of validity is formally limited to $\beta_i \lesssim 0.01$
for our choice of the reduced mass ratio and equal ion and electron background temperatures (Figs.~\ref{pic:Jz_contours}, 
\ref{pic:B_incr}, \ref{pic:B_kurt}, \ref{pic:energy_traces}, and \ref{pic:spectra}).
\end{itemize}

We emphasize that, strictly speaking, our findings are valid only for the particular setup considered and could be modified 
in a more realistic setting. In particular, the results could be modified in a full three-dimensional geometry, which
allows for arbitrary wave propagation angles with respect to the mean magnetic field 
\citep{Howes2008a, Chang2014, Vasquez2014, Howes2015, Wan2015}. Another aspect which might have impacted our results is 
the reduced ion-electron mass ratio of 100. For a realistic mass ratio, the increased separation between ion and electron
scales might improve the agreement between the HK and FK models at {sub-}ion scales as well as potentially reveal certain 
deviations from the phenomenology of KAW turbulence at electron scales \citep{Shaikh2009, Podesta2010}. No definitive answer 
regarding the role of these limitations can be presently given. However, based on a number of previous works, 
it is still reasonable to expect that the simplified setup 
adopted in this work can qualitatively capture at least some of the key features of natural turbulence 
at kinetic-scales of the solar wind \citep{Servidio2015, Li2016, Wan2016}. Furthermore, many of the turbulent 
properties found in this work, such as the dominance of KAW fluctuations \citep{Salem2012, Chen2013}, non-Gaussian statistics at 
kinetic scales \citep{Kiyani2009, Wu2013a, Chen2014, Perrone2016}, and kinetic-scale reconnection 
\citep{Sundkvist2007, Osman2011, Chasapis2015} are supported by in-situ spacecraft observations. 
 
Lastly, apart from exposing the pros and cons of reduced-kinetic treatments, the outcome of the study 
demonstrates that detailed comparisons between fully kinetic and reduced-kinetic models 
can provide significant advantages for elucidating the nature of kinetic-scale dynamics and may also serve as a much 
needed aid for interpreting the results of fully kinetic simulations, as well as guide the 
interpretation of experimental data, acquired from present and (potential) 
future spacecraft missions \citep{Burch2016, Fox2016, vaivads2016}. An ambitious plan to perform a study similar 
to the one presented in this work, but involving an even larger participation from the community, has been 
recently outlined by \citet{Parashar2015a}. For the proposed comparative study, and perhaps others to come, our results
could provide valuable guidelines.  

\acknowledgements
We gratefully acknowledge helpful discussions with O.~Alexandrova, P.~Astfalk, V.~Decyk, T.~G{\" o}rler, Y.~Kawazura, 
W.~Mori, P.~Mu{\~ n}oz, F.~Tsung, and M.~Weidl. 
NFL thanks A.~Schekochihin
and A.~Zocco for useful discussions on the KREHM model.
The research leading to these results 
has received funding from the European Research Council under the European Union's Seventh
Framework Programme (FP7/2007-2013)/ERC Grant Agreement No.~277870. 
C.~Willmott was supported by MIT's Charles E.~Reed Faculty Initiative Fund.
N.~Loureiro was partially supported by the NSF-DOE partnership in basic plasma
science and engineering, award no.~DE-SC0016215, 
and by NSF CAREER award no.~1654168. Furthermore, this work was
facilitated by the Max-Planck/Princeton Center for Plasma Physics.
Computer resources for the hybrid-kinetic and fully kinetic 
simulations have been provided by the Max Planck Computing and Data Facility and by the Gauss Centre for
Supercomputing/Leibniz Supercomputing Centre under grant pr74vi, for which we also acknowledge 
Principal Investigator J.~B{\" u}chner for providing
access to the computing resource. We also acknowledge the Italian 
supercomputing center CINECA (Bologna) under the ISCRA supercomputing initiative.
The gyrokinetic simulations presented in this work used resources of the National Energy 
Research Scientific Computing Center, a DOE Office of Science User Facility supported by the Office of 
Science of the U.S. Department of Energy under Contract No.~DE-AC02-05CH11231. 
Computations with the Viriato code were conducted at Massachusetts Green High
Performance Computing Center on the Plasma Science and Fusion Center partition of
the Engaging cluster, supported by the Department of Energy.
The authors would like to acknowledge the OSIRIS Consortium, 
consisting of UCLA and IST (Lisbon, Portugal) for the use of OSIRIS, for 
providing access to the OSIRIS framework.

\appendix

\section{Numerical details}
\label{app:numerics}

For the fully kinetic, electromagnetic simulations, we use the particle-in-cell (PIC) code \textsc{OSIRIS} \citep{Fonseca2002, Fonseca2008}. 
The reader is referred to \citet{Dawson1983} and \citet{Birdsall2005} for a detailed overview of the PIC method.
Parameters specific to the PIC simulations are given in Table~\ref{tab:osiris}.  A 2nd order compensated binomial filter
is applied on the electric current at each time step and we perform all simulations 
using cubic spline particle shape factors \citep{Birdsall2005}. 
For the problem type considered here, preliminary tests have shown that cubic shape factors deliver superior numerical results, compared to lower-order quadratic and linear splines, due to better energy conservation properties and a significant reduction in the amount of so-called PIC noise. However, even with the use of large 
numbers of particles and smooth particle shapes, the background thermal fluctuations may still mask the 
fluctuations arising from the turbulent cascade which are of main interest here. Therefore, to further reduce PIC noise we also 
employ short-time averages of the turbulent fields. A detailed discussion regarding the time averaging is given
in Appendix~\ref{app:noise}.

\begin{table}[htb!]
\centering
\begin{tabular}{l l l l l l}
\hline
\hline
\multicolumn{6}{c}{\bf FK (PIC)}\\
\hline
Run & $N_x$ & $N_{ppc}$ & $\Delta x/\lambda_D$ & $v_{th,e}/c$ & $\omega_{pe}/\Omega_{ce}$ \\
\hline
A1 & $2048^2$ & 625 & 1.0 & 0.174 & 1.822 \\
A2 & $2048^2$ & 625 & 1.0 & 0.174 & 1.822 \\
B1 & $1920^2$ & 1024 & 1.0 & 0.185 & 3.820 \\
B2 & $1920^2$ & 1024 & 1.0 & 0.185 & 3.820 \\
\hline\hline
\end{tabular}
\caption{Numerical parameters for the FK PIC simulations:
the real space resolution ($N_x$), number of particles per cell per species ($N_{ppc}$), 
the grid spacing in units of the Debye length ($\Delta x/\lambda_D$), 
the electron thermal velocity compared to speed of light ($v_{th_e}/c$), and 
the plasma frequency to cyclotron frequency ratio ($\omega_{pe}/\Omega_{ce}$).\label{tab:osiris}}
\end{table}

The HK simulations are performed with the Eulerian hybrid Vlasov-Maxwell solver \textsc{HVM} 
\citep{Valentini2007}. For a detailed description of numerical algorithms 
employed by the \textsc{HVM} code, we refer the reader to \citet{Matthews1994} and \citet{Mangeney2002}. 
Parameters specific to the HK simulations are given in Table~\ref{tab:hvm}. 
An isothermal equation of state is assumed for the electrons. 
No explicit resistivity is used, but one effectively exists due to the low-pass filters that 
are used by the code's algorithm~\citep{Lele1992}. 
For better consistency with the FK, GK model, and KREHM, 
the HK simulations are performed using the generalized Ohm's law given by Eq.~\eqref{eq:ohm}, which retains 
electron inertia~\citep{Valentini2007}.

\begin{table}[htb!]
\centering
\begin{tabular}{ccccc}
\hline
\hline
\multicolumn{5}{c}{\bf HK (Eulerian)}\\
\hline
Run & $N_x$ & $N_v$ & $\Delta x/d_e$  & $v_{\max}/v_{th}$ \\
\hline
A1 & $512^2$ & $51^3$ & 0.49 & 3.54 \\
A2 & $512^2$ & $51^3$ & 0.49 & 3.54 \\
B1 & $512^2$ & $51^3$ & 0.49 & 3.54 \\
B2 & $512^2$ & $51^3$ & 0.49 & 3.54 \\
\hline\hline
\end{tabular}
\caption{Parameters for the Eulerian HK simulations: the real 
space resolution ($N_x$), the velocity space resolution ($N_v$), grid spacing ($\Delta x/d_e$), 
and the maximally resolved velocity ($v_{\max}/v_{th}$).\label{tab:hvm}}
\end{table}

The GK simulations are carried out with the Eulerian GK code \textsc{Gene} \citep{Jenko2000}. 
Parameters describing the numerical settings for the Eulerian GK simulations are given in Table~\ref{tab:gene}. 
Only one simulation is performed for type A and type B runs because $\epsilon$ does not appear as a parameter 
in the normalized GK equations. For the Orszag-Tang vortex initialization, we 
follow the implementation details described in \citet{Numata2010}. 
Numerical instabilities at the grid scale are eliminated by adding high-order hyperviscous ($\sim k_{\perp}^8$) 
and hypercollisional terms to the perpendicular dynamics in spectral space and to the parallel dynamics 
in velocity space, respectively. With \textsc{Gene} being a grid-based spectral code, 
the (artificial) symmetry of the Orszag-Tang vortex is reflected in the numerical solution 
at all scales with high precision. 
In the FK model, on the other hand, minor asymmetries are naturally present due to PIC noise. 
In order to incorporate slight deviations from the ideal symmetry into the initial condition, we use an approach  
similar to the one employed by \citet{TenBarge2014a} and add a small amount of random perturbations to 
the electron and ion distribution functions. The perturbations are limited to the 
wavenumber range $k_{\perp} \lesssim 1/d_e$.

\begin{table}[htb!]
\centering
\begin{tabular}{lclclclclcl}
\hline
\hline
\multicolumn{6}{c}{\bf GK (Eulerian)}\\
\hline
Run & $N_x$ & $N_{v_{\parallel}}$ & $N_{\mu}$ & $\Delta x/\rho_e$ & $v_{\max,s}/v_{th,s}$ \\
\hline
A & $1536^2$ & 32 & 16 & 0.73 & 3.5 \\
B & $768^2$  & 32 & 16 & 0.65 & 3.5 \\
\hline\hline
\end{tabular}
\caption{Parameters for the Eulerian GK simulations: 
total number of grid points in spectral space ($N_x$), parallel velocity resolution ($N_{v_{\parallel}}$),
$\mu$ space resolution ($N_{\mu}$), grid spacing ($\Delta x/\rho_e$), and the 
maximally resolved species velocity ($v_{\max,s}/v_{th,s}$). Note that the number 
of fully dealiased modes in spectral space is given by $(2/3)^2N_x$. \label{tab:gene}}
\end{table}

For the KREHM simulations we use the \texttt{Viriato} code \citep{Loureiro2016}. The chosen numerical parameters are 
listed in Table~\ref{tab:viriato}. The implementation details for the 
Orszag-Tang vortex are described in the abovementioned reference. Similar to \citet{Loureiro2013}, hyperviscous and 
hyperresistive terms are added to the perpendicular dynamics in order to terminate the turbulent 
cascade before it reaches the limits of the numerical grid. For the parallel velocity space, a model hypercollision operator is adopted. 
Even though KREHM is a low-beta limit of GK, we still perform separate simulations for the $\beta_i=0.1$ and $\beta_i=0.5$ case
by varying the $\rho_i/d_e$ ratio for the two cases.

\begin{table}[htb!]
\centering
\begin{tabular}{cccc}
\hline
\hline
\multicolumn{4}{c}{\bf KREHM}\\
\hline
Run & $N_x$ & $M$ & $\Delta x/d_e$ \\
\hline
A & $1536^2$ & 30 & 0.16 \\
B & $768^2$  & 30 & 0.33 \\
\hline\hline
\end{tabular}
\caption{Parameters for the KREHM Fourier-Hermite spectral simulations:
total grid size in spectral space ($N_x$), size of the Hermite basis for $v_{\parallel}$ ($M$), 
and the grid spacing ($\Delta x/d_e$). \label{tab:viriato}}
\end{table}

\section{Reduction of particle noise by time averaging}
\label{app:noise}

{

In order to reduce the background thermal fluctuations, originating from discrete particle effects in the fully kinetic PIC 
simulations, we average the raw simulation data over a time window of duration $\Delta t \approx 0.5\Omega_{ci}^{-1}$, 
resulting in a sample of 
over 1,000 and 2,000 snapshots for the $\beta_i=0.1$ and $\beta_i=0.5$ simulation runs, respectively.
The time-averaged data is used on an as-needed basis, in cases where there are good reasons 
to believe that some particular result would have been otherwise significantly affected by the particle noise. 
In our present understanding, the relative strength of the background particle noise is 
a highly problem-dependent property, depending on the numerical as well as physical parameters of the PIC 
simulation. Furthermore, the degree to which the particle noise masks the collective, self-consistent plasma response can 
vary even between different types of data of a single simulation. It is in our opinion therefore best to consider various noise reduction
approaches (if at all needed) on a case-by-case basis by visually inspecting the data in real space and by considering 
the spectral properties of the solutions in comparison to the estimated background particle noise. 
 Below we provide a detailed discussion and analysis of the effects of time averaging, comparing the 
time-averaged data with the raw data and with the estimated level of background particle noise. 
We also briefly discuss our results for the out-of-plane electric field, $E_z$, even 
though this quantity is otherwise never used in our comparative study.

In Figure~\ref{pic:noise_xy} we compare the raw data for $E_x$, $E_z$, and $J_z$ with 
the time-averaged data in real space for the $\beta_i=0.1$ set of runs around 4.7 eddy turnover times. The contour plots are 
zoomed into a subdomain of the entire simulation plane to highlight the small-scale structure of the fields.
Qualitatively similar results are obtained for the $\beta_i=0.5$ case (not shown here). The $E_x$ and $J_z$ fields 
have been low-pass filtered to wavenumbers $k_{\perp} < 4/d_e$, which is slightly higher than the 
maximal wavenumber still considered for the comparison of spectral properties between 
different models (the wavenumber range below the gray-shaded regions in Figs.~\ref{pic:spectra} and \ref{pic:spectral_ratios}). 
The $E_z$ field is low-pass filtered to wavenumbers $k_{\perp} < 2/d_e$. 
As Fig.~\ref{pic:noise_xy} evidently shows, the relative strength of PIC noise is different 
for different types of data. Time averaging makes $J_z$ only slightly smoother without significantly altering the shape 
of the turbulent structures, whereas the time-averaged $E_x$ and $E_z$ are significantly less affected by noise than their 
raw data counterparts. For $E_x$, time averaging not only retains any features still visible in the raw data but even reveals 
certain small-scale properties which would otherwise remain hidden in the background noise. For $E_z$, on the other hand, the 
result shows that the raw data is completely dominated by noise even when very aggressively low-pass filtered. 
It is therefore difficult to confidently determine if time averaging for $E_z$ is in our case 
appropriate or not, because the amount of noise in the raw data is too large to clearly recognize any kind of small-scale
turbulent structures which suddenly appear in the time-averaged PIC data. For these reasons, we never use $E_z$ when comparing the 
results obtained from different kinetic models.

\begin{figure*}[htb!]
\epsscale{1.16}
\plottwo{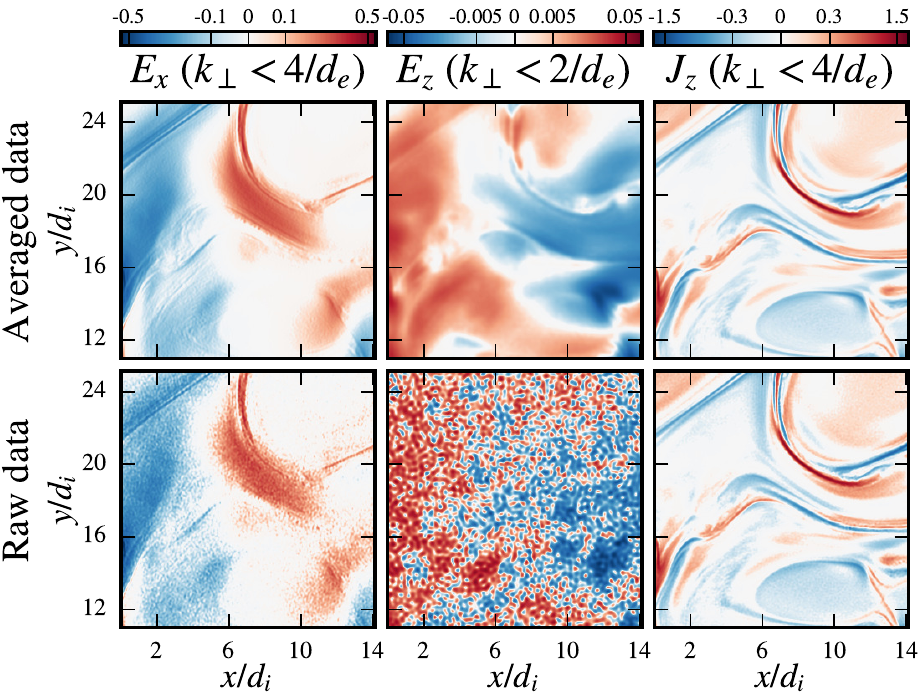}{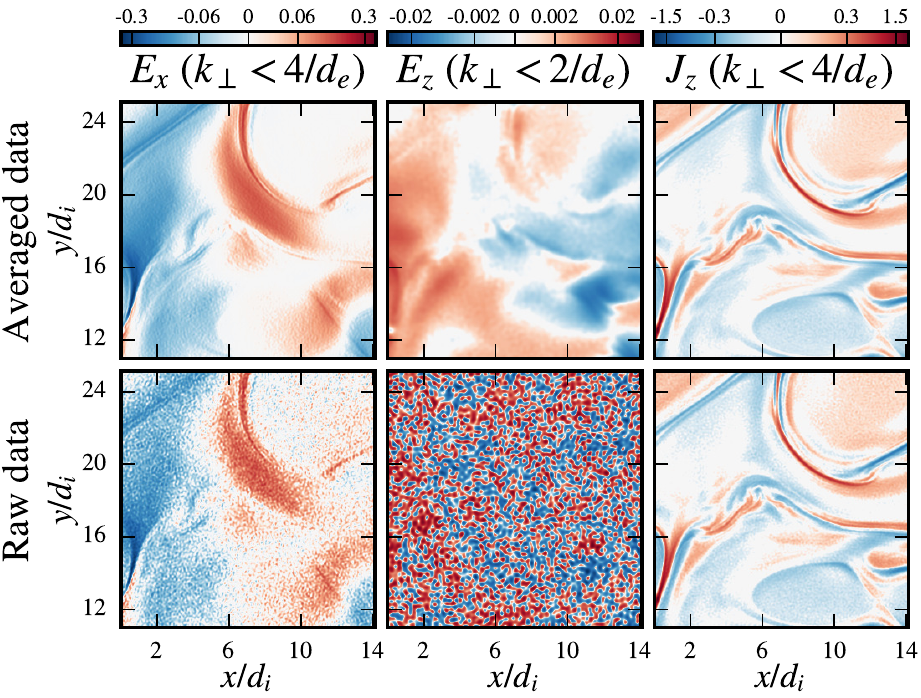}
\caption{\label{pic:noise_xy}  Comparison of the raw and time-averaged PIC data in the $\beta_i=0.1$ simulations 
around 4.7 eddy turnover times for $\epsilon=0.2$ (left) and $\epsilon=0.1$ (right). A doubly-logarithmic scale is used, similar as 
in Fig.~\ref{pic:Jz_contours}. Only a subpart of the entire $x-y$ domain is shown to highlight the small-scale features of the 
turbulent fields.}
\end{figure*}

In Figure~\ref{pic:noise_spec} we compare the spectra calculated from the raw and time-averaged data to the estimates for the 
background particle noise. The turbulent spectra are shown at a single time in the 
simulation at around 4.7 eddy turnover times. All curves are normalized in such a way that 
$\sum_{k_{\perp}}P(k_{\perp})$ gives the mean square value of each field in the default physical 
units adopted in this study (see Sec.~\ref{sec:desc}). The background particle noise is estimated as follows. For each $\beta_i$
we initialize a (uniform) thermal plasma with the exact physical parameters as in the turbulence simulations (including a mean
magnetic field) but without any (smooth) perturbations that would drive the turbulent cascade. We then integrate the 
system using the exact same numerical parameters for a few thousand time steps, until the electromagnetic fluctuations driven by
discrete particle effects attain a quasi-steady state. We then proceed to calculate the spectra from the thermal plasma simulations
and use the results as a proxy for the particle noise in the turbulent runs. 
As such, our estimate neglects the possible influence of turbulence (in particular, spatial and temporal 
variations in plasma parameters) on the background noise. However, given the fact that the 
raw turbulent spectra in Fig.~\ref{pic:noise_spec} are in relatively good agreement with the PIC noise proxy at the largest 
wavenumbers, our estimate still appears to be a reasonable one and we are presently 
unaware of a better alternative for estimating the PIC noise over the entire range of wavenumbers in a turbulent simulation.
As shown in Fig.~\ref{pic:noise_spec}, time averaging significantly modifies the turbulent spectra only 
over the range of wavenumbers, where the raw data is not well-separated from the estimated noise floor (black curves 
in Fig.~\ref{pic:noise_spec}). Light red/orange shading is used here
to indicate the spectral amplitudes that are no more than an order of magnitude below the reference raw data curves.
At each given scale, the raw data is decomposed of a mixture of the turbulent signal and 
particle noise. Therefore, the raw simulation data should be well-separated from the noise floor 
for the (raw) spectrum to be regarded as physically meaningful at some given scale.  
The critical wavenumber at which the time-averaged data deviates from the raw data is generally lower for the simulations 
with a lower value of $\epsilon$ (red/blue curves in Fig.~\ref{pic:noise_spec}). This can be naturally explained by the 
fact that the relative ``signal to noise'' ratio is lower in these simulations, because the particle noise does 
not critically depend on the turbulence amplitude in the first approximation.
Similar to a recent work by \citet{Haggerty2017}, we find that 
time averaging modifies the $E_z$ spectrum at very large scales. However, as shown in Fig.~\ref{pic:noise_spec}, it is worth
considering the fact that the relative strength of particle noise is significantly higher for $E_z$ than for any other field. 
Thus, comparing the turbulent spectra obtained from raw data with those calculated from time-averaged data, without
explicitly estimating the strength of particle noise and without visually comparing the data in real space, 
is not always sufficient to determine whether time 
averaging is appropriate or not. In summary, we conclude that the time averaging over $\Delta t \approx 0.5\Omega_{ci}^{-1}$ does not seem to
significantly modify our simulation results over the range of scales which are well-separated from the noise floor. 
It does, on the other hand, allow for a more detailed investigation of turbulent structures (in particular for $E_{\perp}$), which are not
nearly as clearly recognizable in the raw simulation data.

\begin{figure*}[htb!]
\epsscale{1.}
\plottwo{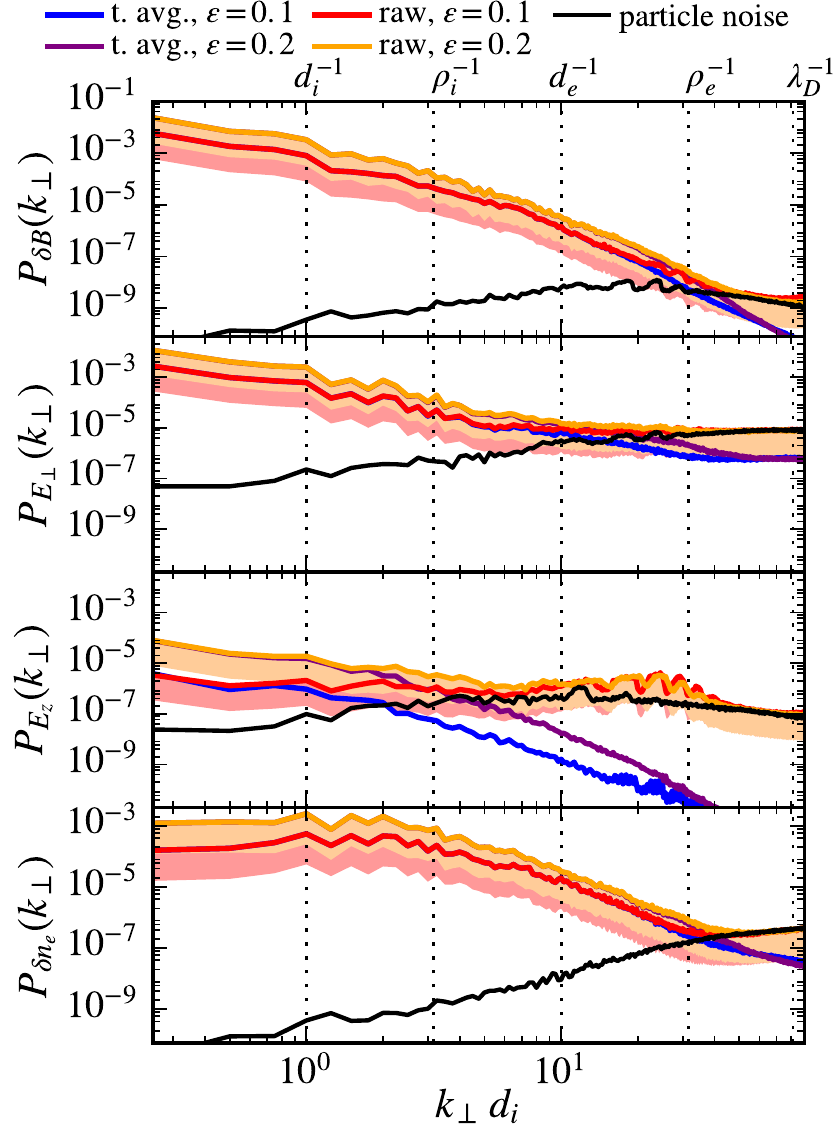}{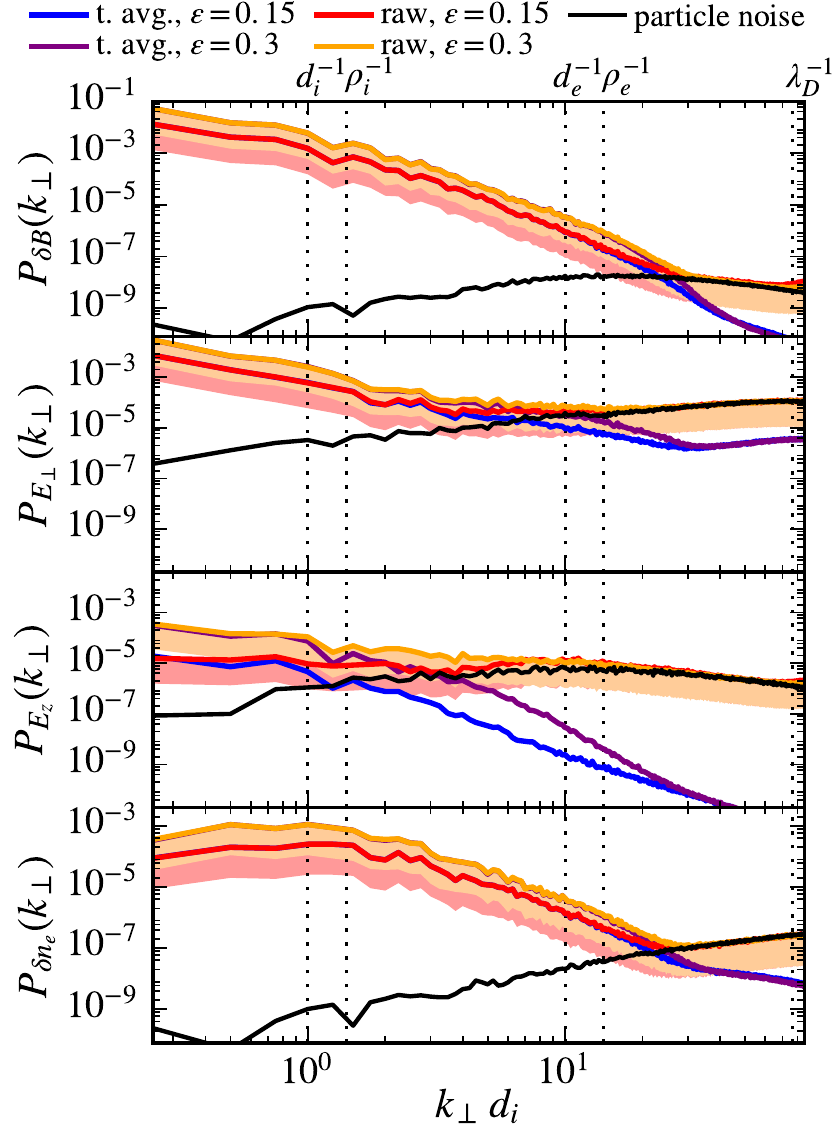}
\caption{Comparison of the turbulent spectra obtained from raw and time-averaged data to the 
estimated background particle noise for $\beta_i=0.1$ (left) and $\beta_i=0.5$ (right). 
Light red (orange) shading is used to indicate the spectral amplitudes that are no more than 
an order of magnitude below the reference raw data curves in the low (high) $\epsilon$ simulations. 
As long as the shading is above the black line, the amplitude of the raw spectrum 
is more than an order of magnitude above the estimated noise floor.\label{pic:noise_spec}}
\end{figure*}

}



\end{document}